\documentclass[prb,twocolumn,aps,showpacs,superscriptaddress,epsfig]{revtex4-2}

\usepackage{amssymb}
\usepackage{amsbsy}
\usepackage{amsmath}
\usepackage{bm}
\usepackage{physics}
\usepackage[usenames,dvipsnames, pdftex, table]{xcolor}
\usepackage{epsfig}
\usepackage[bb=boondox]{mathalfa}
\usepackage{lipsum}
\usepackage{graphicx}
\usepackage{subfigure}
\usepackage{ulem}
\usepackage{dsfont}
\DeclareMathOperator{\diag}{diag}
\usepackage[toc,page]{appendix}
\usepackage{array}
\usepackage{mathrsfs}
\usepackage{soul}
\usepackage{float}
\usepackage{hyperref}
\hypersetup{
	colorlinks=true,
	linkcolor=blue,
	filecolor=cyan,      
	urlcolor=magenta,
	citecolor=red,
}

\begin{document}
\title{ Enhanced Andreev Reflection in Flat-Band Systems: Wave Packet Dynamics, DC Transport and the Josephson Effect}
\author{Sarbajit Mazumdar}
\email{sarbajit.mazumdar@uni-wuerzburg.de}
\affiliation{Institut für Theoretische Physik und Astrophysik und Würzburg-Dresden Cluster of Excellence ct.qmat,  
Universität Würzburg, Am Hubland Campus Süd, Würzburg 97074, Germany
}
\affiliation{National Institute of Science Education and Research, Jatni, Odisha 752050, India}
\affiliation{Homi Bhabha National Institute, Training School Complex, Anushakti Nagar, Mumbai 400094, India }

\author{Anamitra Mukherjee}
\affiliation{National Institute of Science Education and Research, Jatni, Odisha 752050, India}
\affiliation{Homi Bhabha National Institute, Training School Complex, Anushakti Nagar, Mumbai 400094, India }

\author{Kush Saha}
\email{kush.saha@niser.ac.in}
\affiliation{National Institute of Science Education and Research, Jatni, Odisha 752050, India}

\affiliation{Homi Bhabha National Institute, Training School Complex, Anushakti Nagar, Mumbai 400094, India }

\author{Sourin Das}
\email{sdas@isserk.ac.in}
\affiliation{Department of Physical Sciences, IISER Kolkata, Mohanpur, West Bengal 741246, India}

\date{\today}


\begin{abstract}
We investigate Andreev reflection (AR) in a proximity-induced normal-superconductor (NS) junction within the extended \(\alpha-\mathcal{T}_3\) lattice, emphasizing the impact of flat bands on AR. Our findings reveal that flat bands significantly enhance AR. Through wave packet dynamics, we track the real-time evolution of quasi-particle wave packets across the junction, providing deeper insight into electron-hole conversion. Notably, the combination of band flatness and anisotropic dispersion in the $k_x-k_y$ plane induces an electronic analog of Goos-Hänchen (GH) shifts at the NS interface, exhibiting directional asymmetry along the junction. This asymmetry leads to a Hall-like response in Josephson junction in SNS geometry, where transport across the junction region is dominated by the quasi-flat bands.


\end{abstract}
\maketitle

\section{Introduction}
 Andreev reflection (AR) is a fundamental quantum mechanical process occurring at the interface between a normal metal and a superconductor, where an incoming electron from the normal side is  retro-reflected (conventionally) as a hole, leading to the formation of Cooper pairs in the superconductor \cite{RevModPhys.80.1337,PhysRevLett.97.067007,PhysRevLett.102.180405,PhysRevLett.103.237001,PhysRevB.63.104510,PhysRevLett.74.1657,PhysRevB.83.085413,PhysRevB.69.134507}. This mechanism is crucial for understanding charge transport across metal/superconductor interfaces and has significant implications for superconductor-driven quantum electronics\cite{snyman2009bistability,
seja2021quasiclassical,
prada2020andreev,
zazunov2003andreev}. While AR is well understood in conventional superconducting junctions\cite{soori2025crossed,
liu2025nodal,
wang2024intertwined,
vasiakin2025disorder,
nagae2025spin,
sun2023andreev,
yin2019negative}, its behavior in flat-band superconductors has received attention only recently, following the discovery of superconducting bilayer materials exhibiting flat bands\cite{cao2018unconventional,cao2018correlated,balents2020superconductivity}. In these systems,  quasiparticle transport is notably suppressed while coherent pairs are found to be dominant $-$ a fact that is advantageous for applications requiring higher critical temperatures and minimal quasiparticle interference  \cite{PhysRevResearch.3.033117,PhysRevLett.130.216003,aoki2020theoretical}. Additionally, junction currents across flat-band superconductors are shown to depend on quantum metric\cite{li2024flatbandjosephsonjunctions}. A very recent study also discusses how a flat band modifies the critical tunnel current and the presence of non-zero AR at the metal-superconductor interface \cite{virtanen2025superconductingjunctionsflatbands}. Despite these recent developments, the AR and related junction phenomena in flat-band superconductors remain largely unexplored.

In this work, we present a comprehensive theoretical analysis of Andreev reflection (AR) and Josephson transport in an $\alpha$–$\mathcal{T}_3$ lattice system \cite{PhysRevB.103.165429,PhysRevB.106.094503,PhysRevB.95.235432,PhysRevB.98.075422,PhysRevB.99.045420,PhysRevB.99.205429,PhysRevLett.81.5888} exhibiting a tunable quasi-flat band. We find that tuning the band flatness via next-nearest-neighbor hopping enhances AR and leads to nearly perfect electron–hole conversion near the Fermi level. The associated junction conductance becomes largely independent of the flatness parameter in this regime. Moreover, we uncover significant Goos–Hänchen (GH) shifts in the reflected quasiparticle beams, especially under specular AR conditions, with forward shifts characteristic of pseudospin-1 fermions. Real-time wave packet simulations reveal a gradual and coherent electron-to-hole conversion, consistent with near-unit AR probability. Extending to the SNS geometry, our study shows a phase-dependent Josephson current that exhibits a scaling behavior in the presence of quasi-flat band, with the critical current displaying oscillatory but stabilizing trends as a function of junction length. Importantly, the interplay of band flatness and directional asymmetry in the $k_x-k_y$ plane results in finite transverse Josephson currents, suggesting a planar Hall-like response. These findings collectively underscore the utility of flat-band engineered $\alpha$–$\mathcal{T}_3$ systems for realizing enhanced superconducting interfaces and unconventional Josephson effects.

\section{\label{sec:Hamiltonian} Model Hamiltonian}

We begin with the tight-binding Hamiltonian of an $\alpha-\mathcal{T}_3$ lattice with nearest-neighbour (NN) hopping  \cite{PhysRevB.84.241103},
\begin{equation}
   \hat{\mathcal{H}}_{NN} = -t \sum_{\langle ij \rangle} a_{i}^\dagger b_{j} - \alpha\, t \sum_{\langle ij \rangle}  b_{i}^\dagger c_{j} + h.c.,
\label{eq1}
\end{equation}
where $a_{i}$ ($a_{i}^\dagger$) annihilates (creates) an electron at site i on sublattice A (an equivalent definition works for sublattices B and C as well); 
$t$ is the nearest-neighbor hopping between A and B sublattices, $\alpha t$ denotes nearest-neighbor hopping between sublattices B and C. Note that for $\alpha=0$, we recover the graphene lattice, whereas $\alpha=1$ refers to the dice lattice. With the three NN vectors in real space, $\mathbf{r_1}=(\sqrt{3}/2,1/2)a_0$, $\mathbf{r_2}=(-\sqrt{3}/2,1/2) a_0$ and $\mathbf{r_3}=(0,1)a_0$ with $a_0$ being the lattice constant (in nm), the Hamiltonian in momentum-space can be expressed as \cite{PhysRevLett.112.026402,PhysRevB.104.125441,PhysRevB.106.094503}
\begin{equation}
    \hat{\mathcal{H}}_{NN}(\mathbf{k})=\left(\begin{array}{ccc}
0 & f_{\mathbf{k}} \cos \varphi & 0 \\
f_{\mathbf{k}}^{\ast} \cos \varphi & 0 & f_{\mathbf{k}} \sin \varphi \\
0 & f_{\mathbf{k}}^{\ast} \sin \varphi & 0,
\end{array}\right),
\label{eq2}
\end{equation}
where $f_{\mathbf{k}}=-t\left(1+e^{-i \mathbf{k} \cdot \mathbf{r}_{1}}+e^{-i \mathbf{k} \cdot \mathbf{r}_{2}}\right)$, $\mathbf{k}\equiv(k_x,k_y)$ and $\varphi=\tan^{-1}(\alpha)$. The eigenvalues of the Hamiltonian are $\epsilon=0, \pm\left|f_{\mathrm{k}}\right|$. Evidently, the energy dispersion is the same as that of in graphene except for the flat band with energy $\epsilon=0$.  Expanding $f_{\mathbf{k}}$ around the Dirac points $K=\left(\frac{4 \pi}{3 \sqrt{3} a_0}, 0\right)$ and $K'=\left(-\frac{4 \pi}{3 \sqrt{3} a_0}, 0\right)$, we obtain $f_{\mathbf{k}}^{\eta}\simeq\hbar v_F\left(\eta k_x-i k_y\right)$, where $\eta = +(-)$ denotes the $K(K')$ valley and $v_F=-3ta_0$ is the Fermi velocity\cite{PhysRevA.96.033634,PhysRevB.100.085134,PhysRevB.106.245106,Park_2008}.

\begin{figure}
    \centering
    \includegraphics[width =0.5\textwidth]{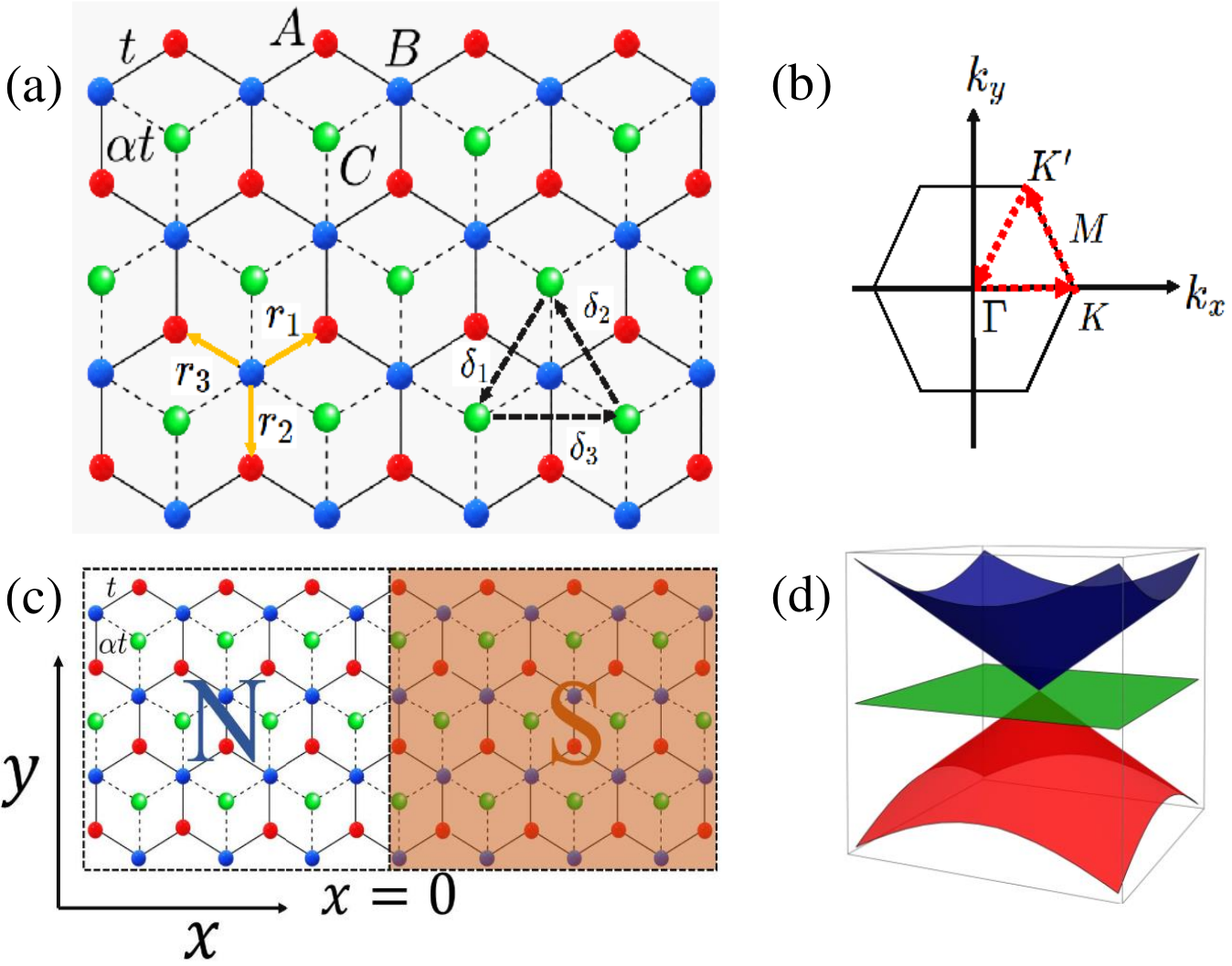}

    \vspace{-1\baselineskip}
    \caption{\textbf{ Schematic for $\alpha-\mathcal{T}_3$ lattice \& junction geometry: }(a) Three sites ($A, B, C$) per unit cell are shown by red, blue and green colour respectively. The NN sites are connected by vectors $r_i$ ($i = 1, 2, 3$) and next-nearest-neighbour sites are connected by vectors $\delta_i$ ($i= 1, 2, 3$), represented by yellow and black (dotted) lines respectively. (b) The right panel shows the path of highly symmetric points in the first Brillouin zone obtained according to the primitive lattice vectors. The dispersion along the highly symmetric part is shown in the Appendix (Fig.~\ref{fig:A1}).
    (c) Schematic for NS junction based on $\alpha-\mathcal{T}_3$ lattice where $x<0$ is a normal metal and $x>0$ is a superconductor. (d) We see the low-energy dispersion around one of the Dirac points. The dispersion is linear and the flat band (for $t_2=0$) is present at $\epsilon=0$, shown in green.}
    \label{fig:1}
\end{figure}

To investigate AR in an NS junction, the Fermi energy on the N side is tuned to lie within the quasi-flat band. To achieve this, we introduce a next-nearest-neighbour (NNN) hopping parameter, $t_2$ for all $A\rightarrow A$, $B\rightarrow B$, and $C\rightarrow C$ hopping in Eq.~(\ref{eq1}) (see Fig.~\ref{fig:1}a). Then the Hamiltonian  takes the form \cite{liu2018generalized,PhysRevResearch.4.033194,PhysRevB.101.235406}
\begin{equation}
    \hat{\mathcal{H}}^{\eta}(\mathbf{k})= \hat{\mathcal{H}}^{\eta}_{NN}+\hat{\mathcal{H}}^{\eta}_{NNN}=\hat{\mathcal{H}}^{\eta}_{NN}+ 2 t_2 \xi(\mathbf{k}) \mathbb{1}_{3\times 3}.
\end{equation}
Here $\mathbb{1}_{3\times 3}$ refers to the identity matrix of dimension 3, $\xi(\mathbf{k})=\sum_{i=1}^{3}e^{i\mathbf{k}.\bm{\delta_i}}$ and $\bm{\delta_i}$ is the NNN vectors as shown in Fig.~\ref{fig:1}a with $\bm{\delta_1}=(\sqrt{3}, 0)a_0$, $\bm{\delta_2}=(-\sqrt{3/2}, 3/2)a_0$ and $\bm{\delta_3}=-(\sqrt{3/2}, 3/2)a_0$. The modified spectrum is found to be  $\epsilon=2t_2\xi(\mathbf{k}),~ 2t_2\xi(\mathbf{k})\pm \left|f_{\mathrm{k}}\right|$. Evidently, $t_2$ makes flat band dispersive and it can allow tuning the degree of flatness.



\section{NS junction and BDG equations}
We consider a plane of $\alpha-\mathcal{T}_3$ lattice in the $(x-y)$ plane. In this setup, the area where $x < 0$ corresponds to the normal metal ($N$) side, while the area where $x > 0$ represents the superconducting ($S$) side, as shown in Fig.~\ref{fig:1}c. The $s$-wave superconducting pair potential $\mathbf{\Delta(x)}=0$ in the normal region and $\mathbf{\Delta(x)}=\Delta_0 e^{i\phi}$ in the superconducting region, via the proximity effect\cite{PhysRevB.104.125441,PhysRevB.101.235417,PhysRevB.106.094503,Zeng_2022}. $\phi$ is the macroscopic phase of the superconducting region. The Dirac-Bogoliubov-de Gennes (DBdG) equation describing
the quasiparticle excitations in the NS junction can be expressed as \cite{PhysRevLett.97.067007,PhysRevLett.74.1657,PhysRevB.108.075425,RevModPhys.80.1337,de2018superconductivity}
\begin{equation}
\mathbb{\hat{H}}
\begin{pmatrix}
u_{e} \\
v_{h}
\end{pmatrix}=
\begin{pmatrix}
\hat{\mathcal{H}}^{\pm}(\mathbf{k}) - E_{\text{F}} & \mathbf{\Delta(x) \mathds{1}} \\
\mathbf{\Delta(x)} \mathds{1}^{\dagger} & E_{\text{F}} - \hat{\mathcal{H}}^{\pm}(\mathbf{k})
\end{pmatrix}
\begin{pmatrix}
u_{e} \\
v_{h}
\end{pmatrix}
= \varepsilon
\begin{pmatrix}
u_{e} \\
v_{h}
\end{pmatrix}.
\label{eq5}
\end{equation}
The component $u_{e}~ (v_{h})$ represents the electron (hole) components of quasi-particle eigenstates, with the energy $\varepsilon$ measured from the Fermi Energy ($E_F$) and $\pm$ refer to valleys $K, K'$. For convenience, we focus only on $\hat{\mathcal{H}}^{+}(\mathbf{k})$ due to valley degeneracy\cite{PhysRevB.104.125441,PhysRevB.108.085423,PhysRevB.106.245106,PhysRevLett.97.067007}. This gives rise to BdG spectra of six bands and the electron-hole dispersion associated with the {\it quasi-flat} is found to be \begin{equation}
\varepsilon(\mathbf{k})=\pm\sqrt{(E_F+U(x)+2 t_2 \xi(\mathbf{k}))^2+\left|\mathbf{\Delta(x)}\right|^2\Theta(x)}.
    \label{eq7}
\end{equation} 
 It is also important to consider a gate voltage potential $U(x) = -U_0\Theta(x)$, $\Theta (x)$ being the Heaviside theta function.

For the reflection at the junction interface (\(x=0\)), the incident energy (\(\varepsilon\)) of an electron from the quasi-flat band and transverse wave vector (\(k_y\)) are conserved. This allows us to express eigenfunctions in the normal and superconducting regions in terms of plane-wave bases with transverse and longitudinal momentum (see Appendix \ref{AppendixA}). Subsequently, we obtain the angle of incidence $\theta=\sin^{-1}[\hbar v_F k_y / (\varepsilon + E_F - 2t_2 \xi(\mathbf{k}))]$ and angle of reflection $\theta'=\sin^{-1}[\hbar v_F k_y / (\varepsilon - E_F + 2t_2 \xi(\mathbf{k}))]$. Accordingly, the critical angle for Andreev reflection (\(\theta_c\)) is found to be \(\theta_c = \sin^{-1}\left[\left|\dfrac{\varepsilon - E_F + 2t_2 \xi(\mathbf{k})}{\varepsilon + E_F - 2t_2 \xi(\mathbf{k})}\right|\right]\), beyond which (\(|\theta| > \theta_c\)) Andreev reflection ceases. For  \(\varepsilon < \Delta_0\), the states in the superconducting region represent coherent superpositions of electron and hole excitations, while for \(\varepsilon > \Delta_0\), they correspond to quasi-electron and quasi-hole states. The probabilities for electron reflection (\(R\)) and Andreev reflection (\(R_A\)) are derived using boundary conditions and the conservation of probability current. Detailed equations and derivations are provided in Appendix \ref{AppendixA}.

\section{Results}\label{sec:results}

\subsection{AR probability and Differential Conductance}

 \begin{figure}
    \centering
  \includegraphics[height=0.25\textheight, width=0.51\textwidth]{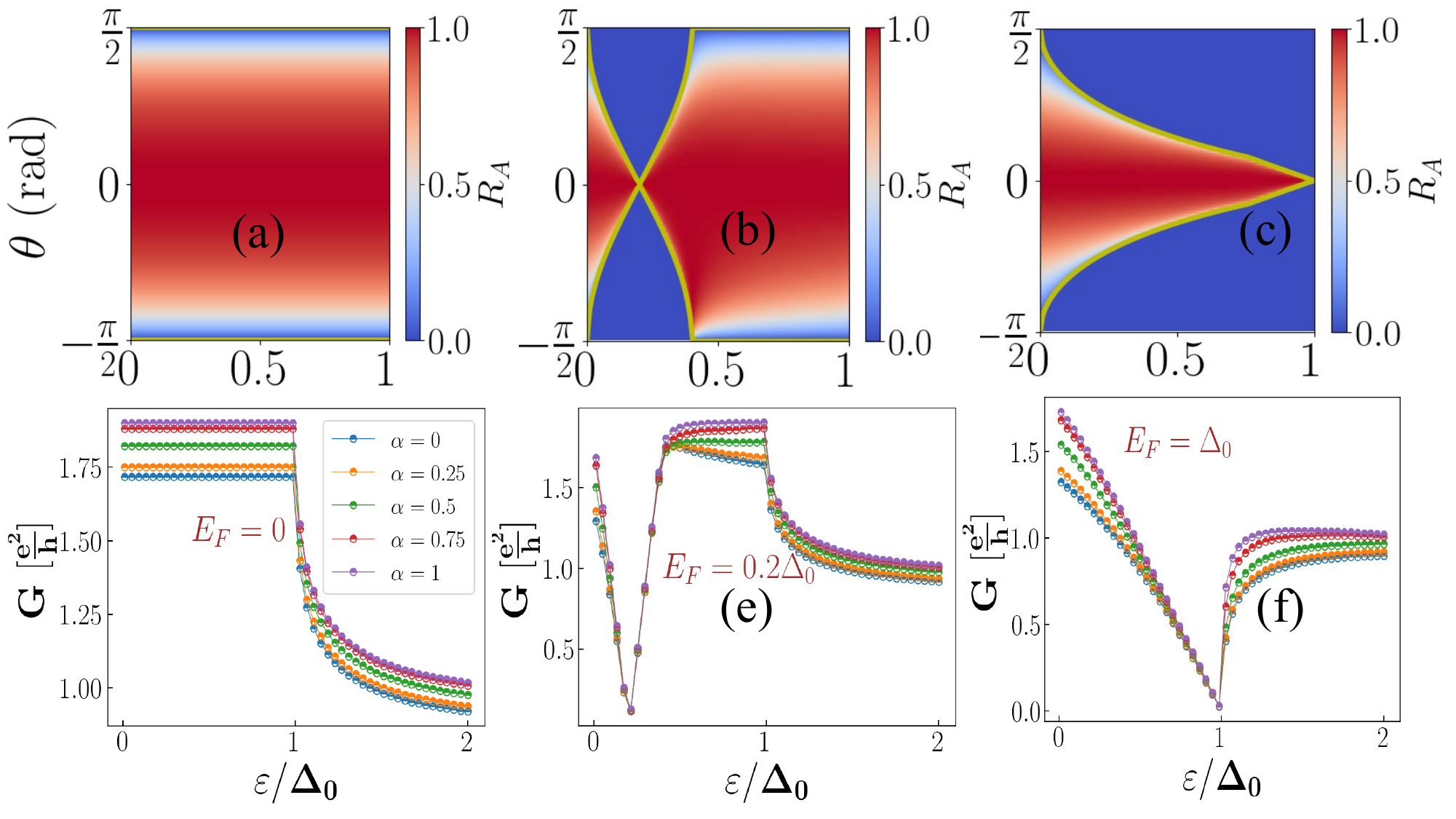}

    \vspace{-1\baselineskip}
    \caption{\textbf{AR probability \& differential conductance }: (a-c) The AR probabilities in $(\varepsilon/\Delta_0-\theta)$ plane for different values of $E_F$ and $\alpha=0.5$. The yellow lines in (b) and (c) denote the locus where $\theta=\theta_c$, representing regions of prohibited AR. The dark blue regions in (b) and (c) represent $R_A \rightarrow 0$, implying electron reflection instead of the hole. (d-f)
    The junction conductance ($G$) as a function of incident energy of electron  for different $E_F$. For $E_F\sim\varepsilon$, the conductance diminishes in (e) and (f), whereas for $\varepsilon > \Delta_0$, $G$ varies weakly with $\alpha$ (d-f). Here we take $\Delta_0=1$ meV, $t_2=0.08$ eV.} 
    \label{fig:2}
\end{figure}  
In the Normal region of the NS junction, the electron reflection probability ($R$) and Andreev reflected probability ($R_A$) obey the probability conservation i.e. $\displaystyle{R+R_A=1}$ for the incoming electron having the energy less than the superconducting gap, $\varepsilon < \Delta_0$. Additionally, the incident angle for the AR of the electron must be less than the critical angle $\theta_c$ ($\left|\theta\right|\lesssim\theta_c$). To illustrate the behavior of  $R_A$ and differential conductance as a function of incident energy $\varepsilon$, we mainly focus on two specific regimes of Fermi energies: (i) $E_F\sim 0$ (specular AR), and (ii) $E_F>0$ (retro AR). 
Note that $\varepsilon\sim\Delta_0$ refers to electron near the flat band.

For specular AR, Fig.~\ref{fig:2}(a) shows that $R_A$ is maximum for a sizable range of incident angle and energy of incident electron at the junction. In contrast, for retro AR, it is forbidden along the yellow line in the $(\varepsilon/\Delta_0-\theta)$ plane as shown in Fig.~\ref{fig:2}(b-c). Moreover, we find regimes of minimum $R_A$ in the  $(\varepsilon/\Delta_0-\theta)$ plane. Note that, in all cases, $R_A$ remains to be finite and maximum when incident energy is pinned at the flat band, i.e, $\varepsilon/\Delta_0=1$.


For AR, a transfer of $2e^{-}$ electronic charge from the normal to the superconducting region across the junction is manifested in the tunnelling conductance ($G$). At zero temperature $G$ is given by \cite{PhysRevB.25.4515}
\begin{equation}
G(\varepsilon)=\frac{e^2}{h}\int~\frac{dk_y}{2\pi}\left(1-R+R_{A}\right),
\end{equation}
where $R$ and $R_A$ both are functions of $\varepsilon$ and $k_y$. Note that $G(\varepsilon)\leq2$(in units of
 $\frac{e^2}{h}$) due to the conservation of probability. In Fig.~\ref{fig:2}(d), we see that for (i) $E_F\sim 0$ (specular AR) and $0<\varepsilon<\Delta_0$, the conductance remains constant across various values of $\alpha$. This is attributed to the maximum $R_A$ arising from evanescent modes for all incident angle $-\pi/2<\theta<\pi/2$. However, as (ii) $E_F$ increases (retro AR), slight variations in conductance profiles are observed for different values of $\alpha$(see Fig.~\ref{fig:2}e). Furthermore, for $\varepsilon\sim E_F=\Delta_0$ (Fig.~\ref{fig:2}f), the conductance diminishes significantly as the $R_A$ is relatively smaller than other cases for a range of incident angle and incident energy. 
 %
 These observations are in {\it stark} contrast with previous findings in conduction band calculations, where variations in $\alpha$ resulted in different conductance profiles and a perfect AR was only achieved for $\varepsilon=\Delta_0$ \cite{PhysRevB.104.125441}. We note that for numerical analysis, we focus on a specific range of Fermi energy which is aligned with the structure of the system's density of states (DOS) and bandwidth (details in Appendix \ref{appenB}).

\vspace{-1mm}
\subsection{Goos–Hänchen shifts}
It is widely known in optics that when a light beam is fully reflected at the interface of two dielectric media, it experiences a lateral displacement from its original incident position. This phenomenon is known as the Goos–Hänchen (GH) shift \cite{Goos1947EinNU}. 
An analogous lateral shift has also been recently observed in the reflected quasiparticle (hole) beam during the Andreev reflection process at junctions, exhibiting distinct characteristics for both pseudospin-$\frac{1}{2}$ and pseudospin-$1$ fermions\cite{PhysRevB.101.235417,PhysRevB.109.035432,PhysRevB.98.075151,PhysRevB.105.085415}.   

To numerically simulate GH shifts in our setup, we consider Gaussian wave packets to represent both electron-like and hole-like beams. Following Refs.~\onlinecite{PhysRevB.101.235417, PhysRevB.105.085415}, we directly compute the GH shift from the reflection amplitudes 
\begin{equation}
    \delta y_{e(h)}=-\frac{\partial \arg[r(r_A)]}{\partial k_y},
    \label{13}
\end{equation}
where $\delta y_{e(h)}$ is the lateral GH shift in the $y$-direction for electron and hole reflections respectively, for an incident electronic beam centered around wave vector ($k_x,k_y$). Here, $r$ and $r_A$ are electron reflection and AR amplitude, respectively. 

\begin{figure}
    \centering
    \includegraphics[width =0.5\textwidth]{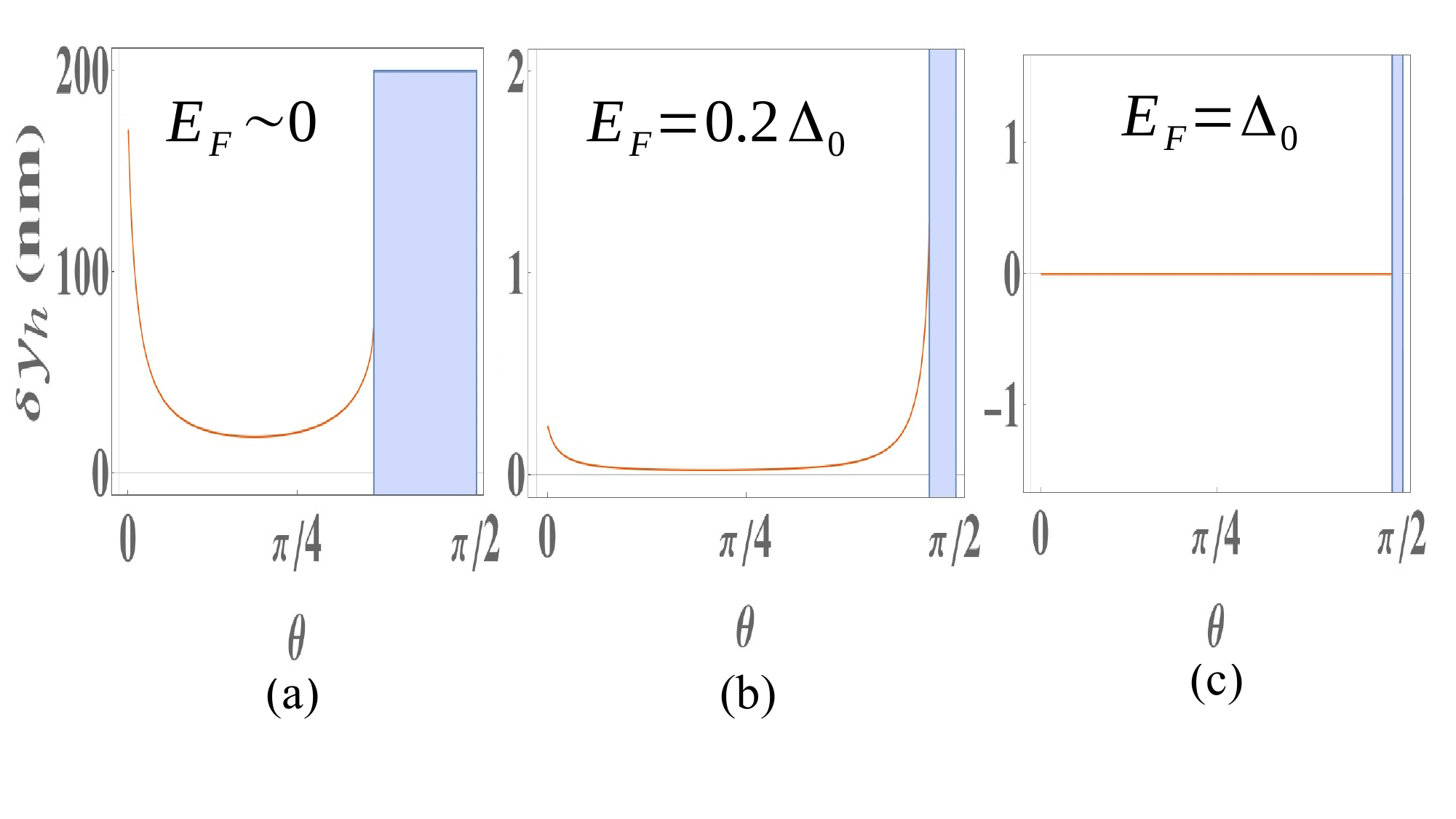}
    \vspace{-1\baselineskip}
    \caption{\textbf{ Goos–Hänchen shifts:} (a-c) The GH shift $\delta y_h$ of the hole-reflected beams as a function of incident angle $\theta$ for a fixed incident energy $\varepsilon=0.99\Delta_0$. The shaded blue area indicates regions where $\theta>\theta_c$, denoting the absence of AR. 
    For $E_F\sim 0$, we find enhanced GH shifts (a) in contrast to (b) and (c) where reduced  shift are found for $E_F=0.2\Delta_0$ and $E_F=\Delta_0$ respectively. Clearly, for retro reflection, the GH shift is much lower than the specular one (a). The parameters are chosen to be $\Delta_0=1$ meV, $\alpha=0.5$, and $t_2=0.08$ eV.}
    \label{fig:3}
\end{figure}
 \begin{figure}
	\centering
	\subfigure []
	{\includegraphics[width=0.5\textwidth]{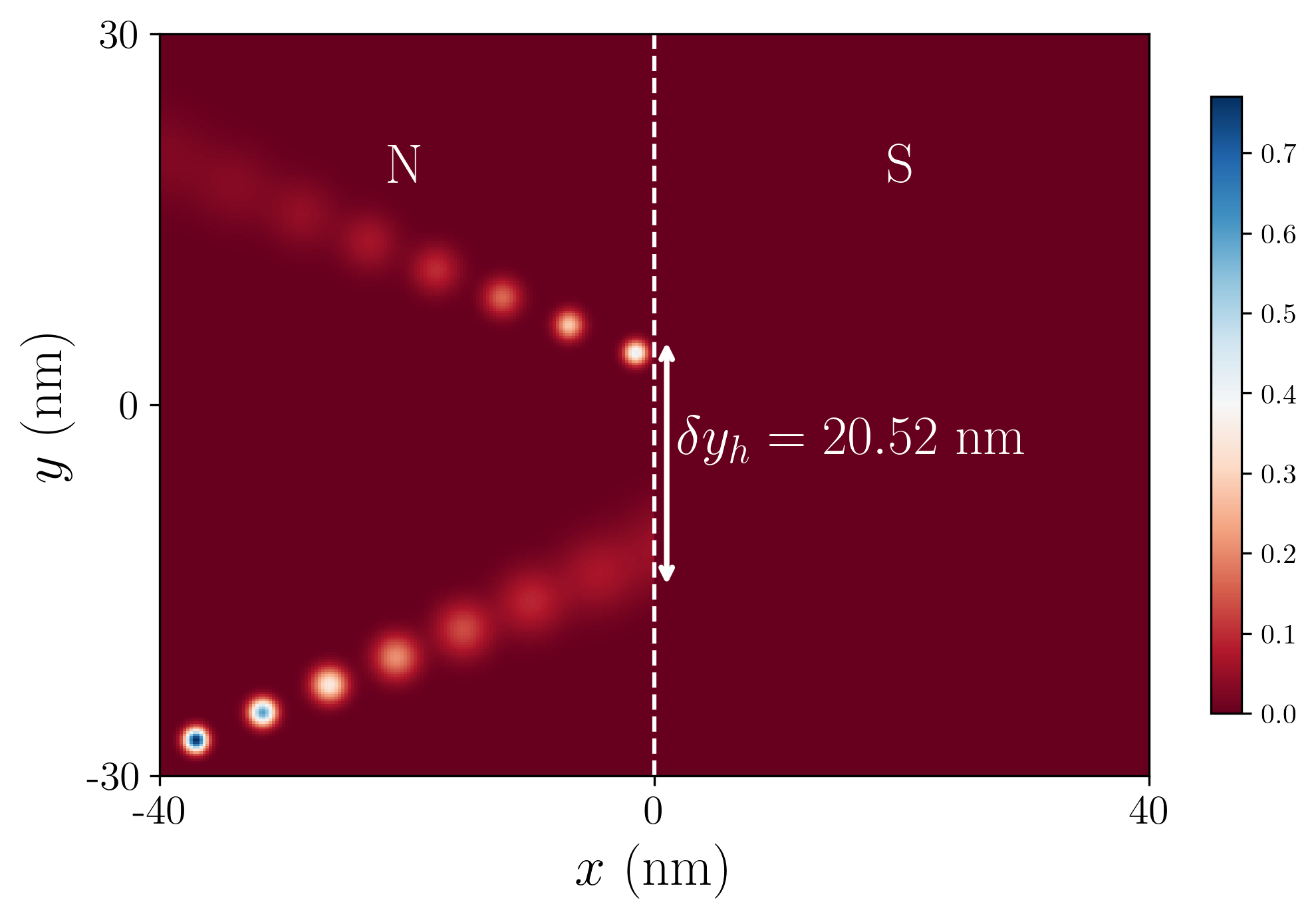}\label{retro}}
	\subfigure []
	{\includegraphics[width=0.45\textwidth]{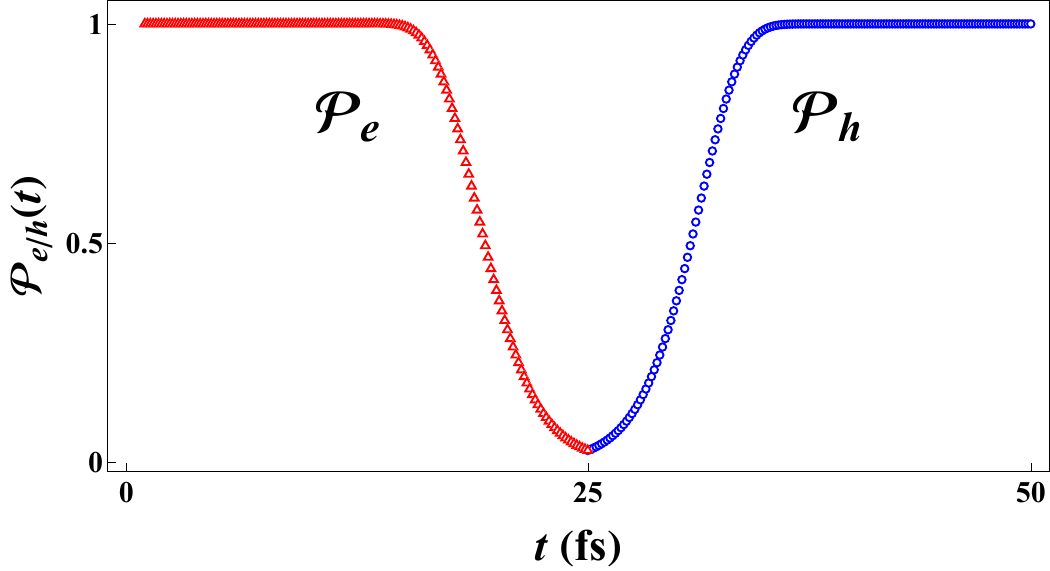}\label{specular}}
	\vspace{-1\baselineskip}
	\caption{\textbf{Wave packet propagation \& Probability distribution:} (a) The incident (reflected) electron (hole) wave packet propagation in the normal region followed by AR for $E_F=0$. Both electronic (hole) wave packets are undergoing dispersion as they move towards (away from) the junction. The initial position of the wave packet is in $(x_0,y_0)\equiv (-37.2,-27.2)~$nm. Other parameters are $\sigma=0.45,~\alpha =0.5,~ t_2=0.08$ eV and $\Delta_0=1$ meV. (b) The plot shows the probability densities ($\mathcal{P}_{e}, \mathcal{P}_{h}$) of the electronic and hole wave packets plotted in red and blue markers respectively. Initially, $\mathcal{P}_{e}$ is equal to 1, indicating that the initial wave packet is normalized over space. However, as time progresses, the electronic wave packet approaches the junction, resulting in the decrease of $\mathcal{P}_{e}$, while $\mathcal{P}_{h}$ increases from zero and eventually reaches $\mathcal{P}_{h}=1$. }.
    \label{fig:4}
\end{figure}

Fig.~\ref{fig:3}(a-c) illustrates the typical GH shifts for various incident electron angles and fixed incident energy $\varepsilon$. We set $\varepsilon=0.99\Delta_0$ to take
advantage of the pronounced density of states due to the presence of quasi-flat band near the gap edge. In Fig.~\ref{fig:3}(a), we observe significant GH shifts as the system undergoes specular AR ($E_F = 0$). This is attributed to the resonance condition being met over a wide range of incidence angles \cite{PhysRevB.110.035420} and directional asymmetry of the Fermi contours at the NS interface along the transverse  momentum ($k_y$) (see Fig.~\ref{fig:8}). Notably, the shift persists up to the critical angle for AR (\(\theta_c\)); beyond this angle, no spatial shift occurs in the AR-forbidden region (indicated in blue). In contrast, when the system enters in the retro-reflection dominated regime ($E_F>0$), the magnitude of the GH shift is considerably smaller as shown in Figs.~\ref{fig:3}(b, c). It is crucial to highlight that for pseudospin-$1$ fermions, the GH shifts typically occur in the forward direction, indicating positive values, unlike the behavior observed for pseudospin-$\frac{1}{2}$ fermions\cite{PhysRevB.101.235417}. This fundamental characteristic remains consistent in our plots and  the sizable GH shifts due to quasi-flat bands suggest promising experimental applications in SNS-type waveguides, optics, acoustics and junction-like systems \cite {PhysRevA.111.033702,
hong2025goos,
villegas2024goos,
waseem2024magnomechanically}. 

\subsection{Wave packet dynamics}
Understanding quasi-particle behavior on a real-time scale is crucial for gaining insights into the system's dynamics, which in turn also validates our analysis in the preceding sections. 
To study the dynamics, we use an initial Gaussian wave packet of incident electrons with a fixed energy. The total wave function in the Normal region can be written 
\begin{equation}
    \Phi=\psi_e^{+}+r\psi_e^{-}+r_A \psi_h^{-},
    \label{eq14}
\end{equation}
where $\psi_{e}^{\pm}$ and $\psi_{h}^{\pm}$ refer to  electron and hole wavefuctions, respectively; the superscripts $\pm$ denote the direction of propagation towards the junction. We consider an initial wave packet in the position space at $t=0$ in the form $
    \Phi(\mathbf{r}, t=0)\sim (u_e ~ v_h)^T\times \mathscr{G}(\mathbf{r}, 0).
\label{eq15}
$, where
 $\mathscr{G}(\mathbf{r}, 0)$ is a Gaussian envelope function expressed in  a plane-wave basis as
\begin{equation}
\mathscr{G}(\mathbf{r}, 0)=\frac{1}{(2\pi\sigma^2)^{\frac{1}{4}}}\exp{ \frac{(x-x_0)^2+(y-y_0)^2}{(2\sigma)^2}+i \mathbf{k_0}\cdot \mathbf{r}}.
    \label{eq16}
\end{equation}
With this, the time-evolved state $\Phi(\mathbf{r}, t)$ is generated by solving numerically $i\frac{d \ket{\Phi}}{dt}= \mathbb{\hat{H}} \ket{\Phi}$ using Explicit Runge-Kutta method of order 5(4) using SciPy library choosing absolute tolerance of $\sim10^{-6}$\cite{virtanen2020scipy}. 
The energy of the wave packet is set to $\varepsilon=0.99\Delta_0$ as before and the incident electron approaches at an angle of $\theta=\pi/4$. We subsequently set $E_F=0$ to observe the specular AR. The dynamic evolution is crucial for accurately capturing real-time behavior, especially given that electrons in the quasi-flat bands are almost stationary. Fig.~\ref{fig:4}(a) illustrates the evolution of localized Gaussian wave packet in the normal region,    
indicating wavepackets for incident electron ($\psi_e^{+}$) and reflected hole ($\psi_h^{-}$). It is worth pointing out that there is typically a reflected electron-wave packet as well. However, we opt not to include it in the visualization because, in cases of near-perfect AR, the electron reflection amplitude $R\rightarrow 0$ and AR amplitude $R_A\rightarrow 1$. 

Fig.~\ref{fig:4}(a) further shows a shift between incident electron wave packet and reflected hole wave packet at the junction and this is referred as the Goos–Hänchen shift, as indicated in the figure by $\delta y_h$.  
 The time evolution also captures how the probability densities ($\mathcal{P}_{e}, \mathcal{P}_{h}$) of the electronic and hole wave packets vary with time, $ \mathcal{P}_{e}(t)=\int \left|\psi_e^{+}(\mathbf{r},t)\right|^2 d\mathbf{r},~\mathcal{P}_{h}(t)=\int \left|\psi_h^{-}(\mathbf{r},t)\right|^2 d\mathbf{r}$, where the integration is performed over the normal region for both cases. In Fig.~\ref{fig:4}(b), we plot how the probability density for each incident (reflected) electron (hole) changes with time; the integrated probability is close to one as the initial wave function is normalised. When the electron wave packet is close to the junction interface, we see that the probability of electronic wave packet $\mathcal{P}_{e}$ decreases and then during the hole reflection, $\mathcal{P}_{h}$ increases and retain the value one at a later time due to finite quasi-particle lifetime \cite{lee2012stickynormalsuperconductorinterface,olivares2014dynamics}. This feature dictates that electron-to-hole conversion during AR is not an instantaneous process in finite-size samples.


\begin{figure}
\includegraphics[width =0.5\textwidth]{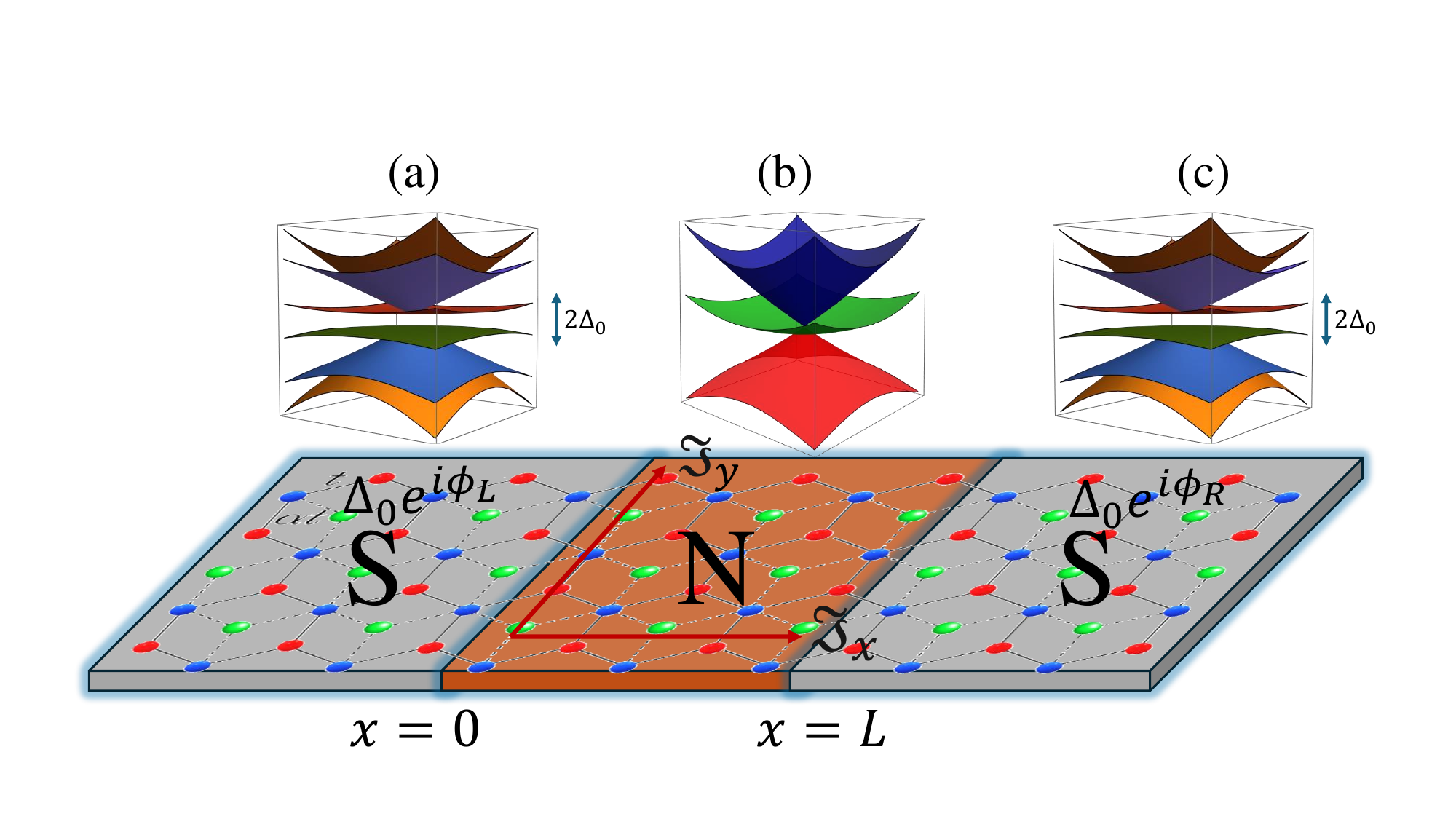}
    \vspace{-1\baselineskip}
    \caption{\textbf{SNS junction:} Schematic of a quasi-flat band-based SNS junction, where a metallic region (located between $x = 0$ and $x = L$) is sandwiched between two superconducting leads characterised by macroscopic phases $\phi_L$ and $\phi_R$, respectively. The upper panels (a–c) display the corresponding band structures for each region, with the bands in the superconducting regions representing BdG bands.}
    \label{fig:5}
\end{figure}

\subsection{SNS junction}
\label{sns}

We now analyze the properties of Josephson current across a flat-band-based SNS junction, where a metallic part ($x=0$ and $x=L$) is sandwiched between two superconducting junctions with macroscopic phases $\phi_L$ and $\phi_R$ respectively as shown in Fig.~\ref{fig:5}. We focus on the short junction regime where the length $L$ of the normal region is less than the superconducting coherence length ($\xi=\frac{\hbar v_F}{\Delta_0}$)\cite{discussion}. Note that such a short junction is mostly applicable to various experimental conditions\cite{PhysRevLett.120.077701,Lee2015UltimatelySB,Pellegrino_2020,PhysRevLett.10.230,matsuo2023josephson}, .
With this, we first introduce the eigenstates in the three regions: in the left ($x<0$) and right ($x>L$) superconductors, the eigenstates are denoted as $\psi_{\mathrm{S}}^{l\pm}$ and $\psi_{\mathrm{S}}^{r\pm}$, respectively; in the normal region ($0<x<L$), the electron- and hole-like eigenstates are denoted by $\psi_{e}^{\pm}$ and $\psi_{h}^{\pm}$, respectively\cite{PhysRevB.104.125441,PhysRevB.102.045132}.
The total wave functions in the superconducting region and the normal region for three distinct regimes of the junction are given by 

\begin{align}
\Psi_{mid} & =\mathcal{A}\psi_{e}^{+}+\mathcal{B} \psi_{e}^{-}+\mathcal{C} \psi_{h}^{+}+\mathcal{D}\psi_{h}^{-} ,\\
\Psi_{left} & =\mathcal{E}\psi_{\mathrm{S}}^{l-}+\mathcal{F} \psi_{\mathrm{S}}^{l+}, \\
\Psi_{right} & =\mathcal{Q}\psi_{\mathrm{S}}^{r-}+\mathcal{M} \psi_{\mathrm{S}}^{r+}. 
\end{align}
\begin{figure}
    \centering
    \includegraphics[width =0.45\textwidth]{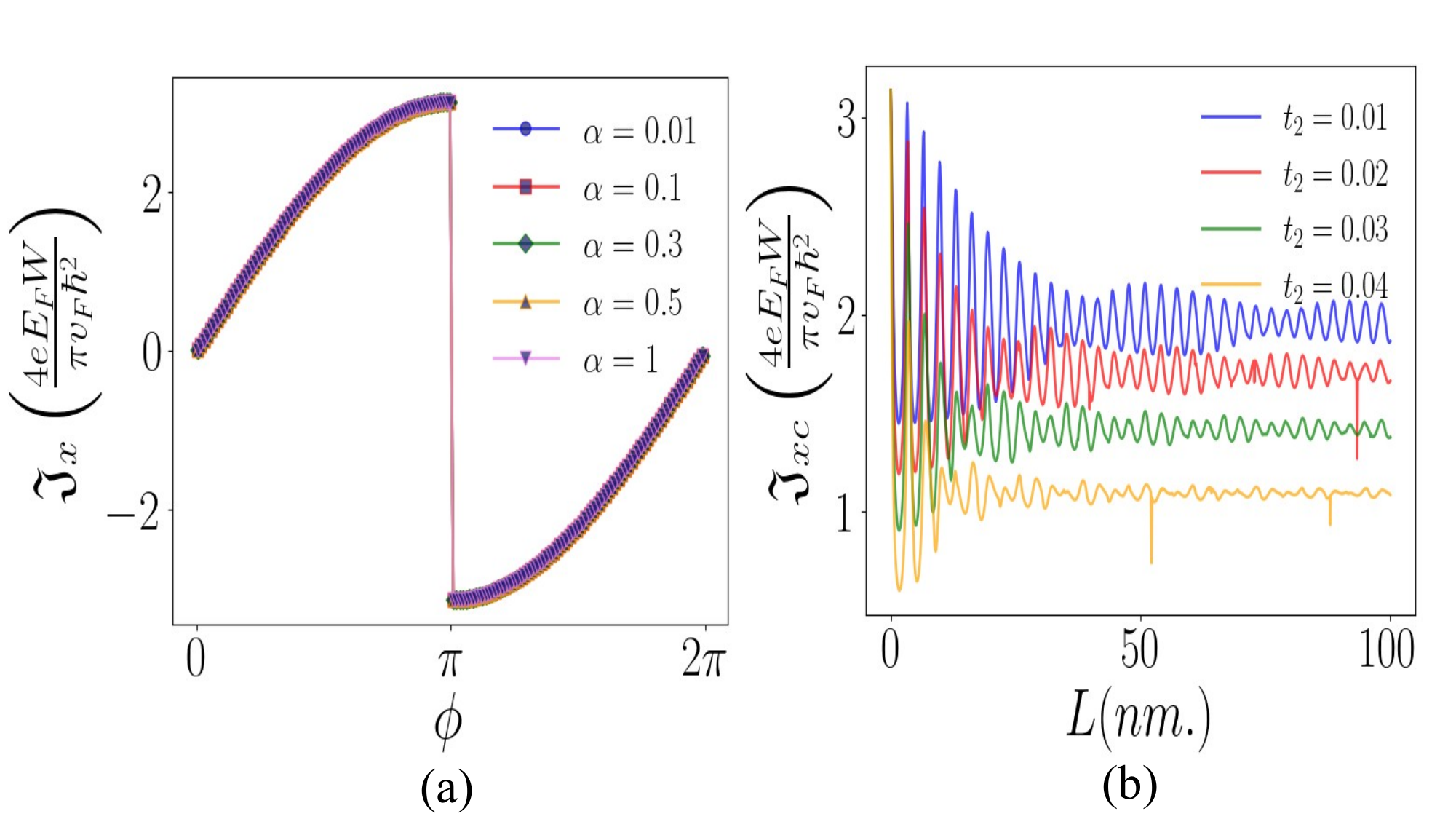}
    \vspace{-1\baselineskip}
    \caption{\textbf{ Josephson current:} (a) The dc-Josephson current for ABS with the macroscopic phase difference. A scaling behavior with parameter $\alpha$ is observed, and the overall nature of the plot is sinusoidal. The parameters are taken to be $E_F=\Delta_0$, $L=10$ nm and $t_2=0.08$. (b) The critical dc-Josephson current as a function of junction length $L$.  Notice that a stabilizing behavior in the critical current as the junction length increases. All plots show an oscillatory nature in general for different flatness parameter $t_2$.}
    \label{fig:6}
\end{figure}
Here, $\Psi_{mid}$ describes the wavefunction in the normal region, while $\Psi_{left}$ and $\Psi_{right}$ describe the left and right superconducting regions. The eight unknown coefficients (\(\mathcal{A}, \mathcal{B}, \mathcal{C}, \mathcal{D}, \mathcal{E}, \mathcal{F}, \mathcal{Q}\), and \(\mathcal{M}\)) are determined by imposing boundary conditions at the two interfaces of the junction, located at \(x=0\) and \(x=L\). Matching the wavefunctions and their derivatives at these boundaries, we obtain eight coupled linear equations. They can be compactly expressed in matrix form as: $\mathbb{A} \mathbb{X} = \mathbb{0}$, where \(\mathbb{A}\) is an \(8\times8\) coefficient matrix depending on the system parameters, and \(\mathbb{X}\) is the vector containing the unknown coefficients. 

For simplicity, we focus on the retro-reflection regime. In this regime, electron-like excitations can be retro-reflected as hole-like excitations, enabling the formation of Andreev bound states (ABS) within the normal region. To admit non-trivial solutions (i.e. \(\mathbb{X} \neq 0\)), the determinant of the coefficient matrix must vanish, i.e. $\det \mathbb{A} = 0$. This yields a transcendental equation for the ABS energies, which we write as
\begin{equation}
    \mathscr{F}(\varepsilon, E_F, L, \theta, \phi, \alpha) = 0,
    \label{eq21}
\end{equation}
where \(\mathscr{F}\) is a function carrying the dependence on \(\varepsilon\), \(E_F\), the junction length \(L\), the angle of incidence \(\theta\) and the superconducting phase difference between the macroscopic phases of two superconducting junctions $\phi=\phi_R-\phi_L$.
Solving Eq.~\ref{eq21} numerically, we find the values of $\varepsilon_{\text{ABS}}$,   keeping the other parameters fixed. The ABS spectra and the analytical calculations are discussed in the Appendix~\ref {appenC}. Using $\varepsilon_{\rm ABS}$, the Josephson current passing through the normal region can be expressed as \cite{PhysRevB.104.125441,PhysRevB.74.041401,PhysRevB.95.064511,PhysRevLett.66.3056}
\begin{equation}
    \mathfrak{J}_x=\mathfrak{J}_0\int \frac{d \varepsilon}{d \phi} \cos \theta~ d \theta.
    \label{eq22}
\end{equation}
Here the pre-factor $\mathfrak{J}_0=-\frac{4 e W E_{\mathrm{F}}}{\pi \hbar^2 v_{\mathrm{F}}}$ accounts for both the valley and spin degeneracy, and $W$ is the transverse width of the sample. In Fig.~\ref{fig:6}a, we plot the phase-dependent Josephson current for various values of $\alpha$. Interestingly, we observe that the phase-dependent current does not change with $\alpha$, which implies a \textit{scaling behavior} in the Josephson current induced in the presence of the quasi-flat band. This is in contrast to the case, when the Fermi energy is away from the quasi-flat band, there is no scaling behavior found in the Josephson current \cite{PhysRevB.104.125441}.
We also observe that the landscape of the Josephson current is sinusoidal and there is a discontinuous jump at $\phi=\pi$ \cite{chi2023electronic,Xu_2017,cheng2008josephson, PhysRevB.62.648,Zhu_2001}. 

To understand how the presence of the flat band influences superconducting transport, we further analyze the behavior of the critical Josephson current \(\mathfrak{J}_{xc}\). The critical current is determined by maximizing \(\frac{d\varepsilon}{d\phi}\). We find that critical current exhibits an oscillatory behavior as a function of \(L\) for the short junction limit due to the transmission resonance phenomenon \cite{PhysRevB.76.054513, madsen2017josephson,PhysRevB.94.094514}. In addition, the mean value of the critical current is enhanced when the flat band tuning parameter \(t_2\) is reduced (Fig.~\ref{fig:6}b), and as the junction length increases, the critical current stabilizes instead of rapidly decaying. Note that, for $L\rightarrow0$, the value of $\mathfrak{J}_{xc}$ is independent of $t_2$.  Hence, these features of the Josephson current show that tuning the flat-band parameter not only modifies the mean critical current but also enhances the stability over increasing junction length, providing an additional degree of control over Josephson transport beyond conventional dispersive band systems.  



\begin{figure}
    
    \subfigure[]{
        \includegraphics[width=4cm,height=4cm]{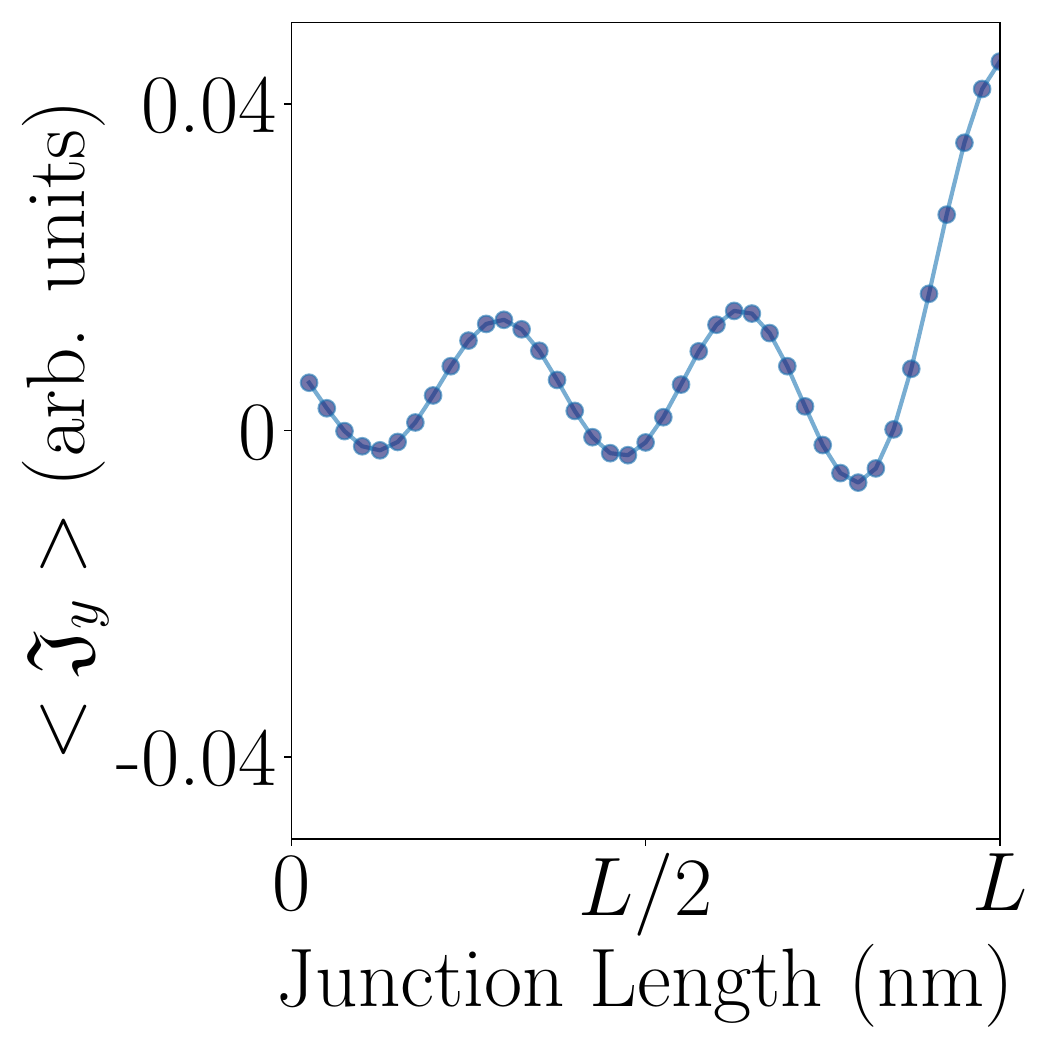}
        \label{fig:wavefunction_edge_16_0.2}
    }
    \subfigure[]{
        \includegraphics[width=4.1cm,height=4.1cm]{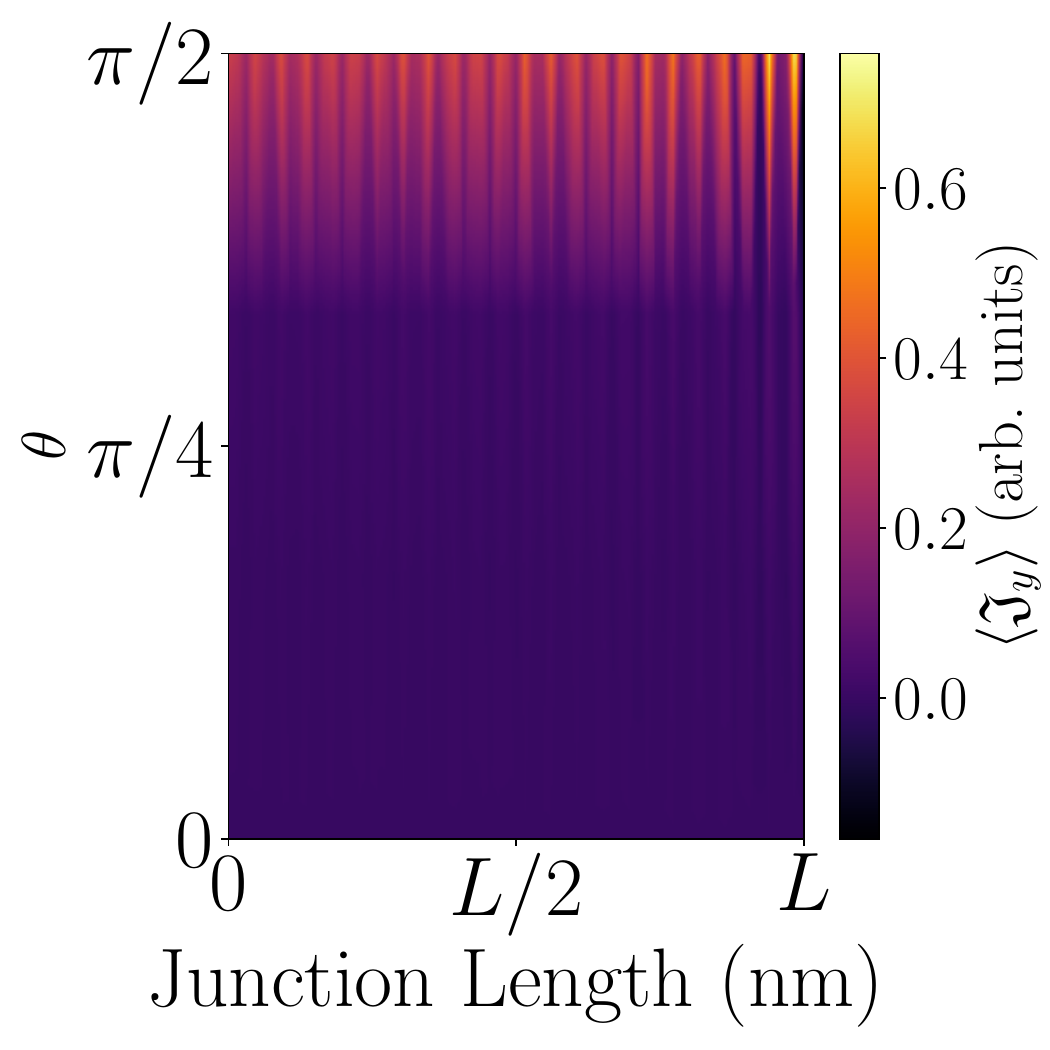}
        \label{fig:wavefunction_edge_16_0.4}
    }
    \caption{\textbf{Transverse current (\(\langle \mathfrak{J}_y \rangle\)) profiles}: (a) The total current contribution (in arb. units) along the \(y\)-direction (the propagation direction of the Andreev modes) shows an oscillatory behavior as the junction length increases. For the plot, the parameters are set as \(\theta = \pi/4\), \(L = 20~\mathrm{nm}\), $W=30~nm$, and \(\phi = \pi/2\).; (b) The total current contribution (in arb. units) along the \(y\)-direction is plotted concerning the incidence angle($\theta)$ and the junction length ($L$).}
    \label{fig:7}
\end{figure}

\begin{figure}
\includegraphics[width =0.52\textwidth]{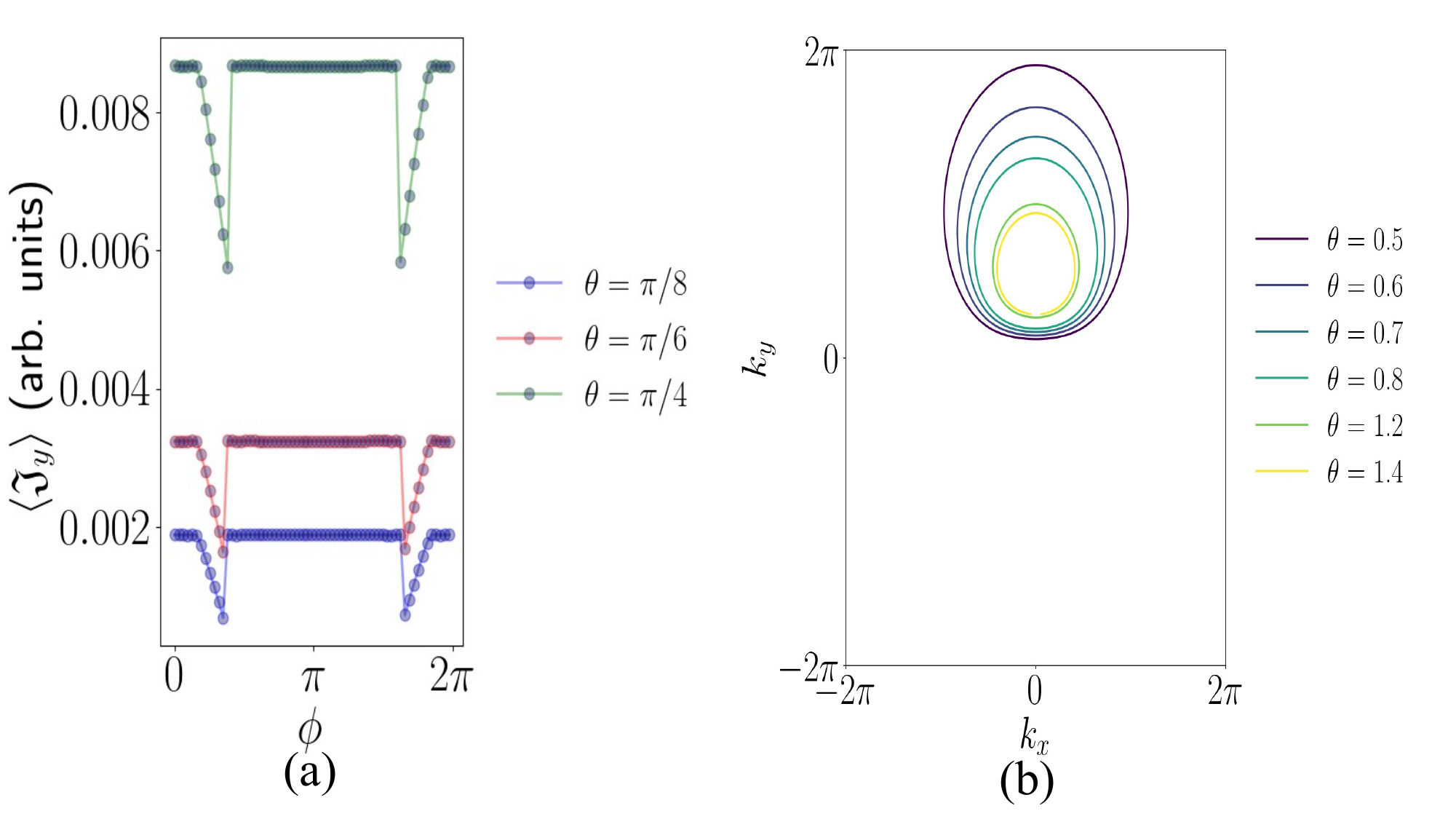}
    \vspace{-1\baselineskip}
    \caption{\textbf{Phase dependent Transverse current \&  Fermi contours:} (a) The transverse charge current as a function of the macroscopic phase difference for different incident angles of electrons. The parameters are $L=20~nm$, $W=30~nm$ and $E_F=0$. For all angles, the transverse current shows Hall-like response. (b) Fermi contours for different incident angles of electrons for $E_F=0$ and $t_2=0.08$, Notably, the contours are asymmetric in $k_y$ direction.}
    \label{fig:8}
\end{figure}

\subsection{Transverse charge current due to Andreev modes}

We investigate the transverse charge current ($y$-direction) in an SNS junction, focusing on the contributions from Andreev modes. The presence of band flatness and anisotropic dispersion in the $k_x-k_y$ plane induces a directional asymmetry at the junction interface. In Fig.~\ref{fig:8}, the Fermi contours for different incident angles are plotted. We see that the contours are not symmetric around the $k_y$ axis. As a result, a finite transverse current arises from the propagating modes in the $y$-direction of the SNS geometry during specular Andreev reflection (AR). According to the Landauer-Büttiker formalism\cite{PhysRevLett.57.1761}, the magnitudes of this transverse charge current involving propagating Andreev modes is given by \( \langle \mathfrak{J}_y \rangle = e \, \mathcal{N}_W \langle \hat{v}_y \rangle \), where \( \hat{v}_y \) is the velocity matrix element, $e$ is the electron's charge and $\mathcal{N}_W$ is the density of transverse modes with transmission width \( W \). 
The density of Andreev modes with energy $\varepsilon\sim \Delta_0$ (flat band contribution) is obtained to be  
\begin{align}
&\mathcal{N}_W = W \int \frac{dk_y}{2\pi} ~\Theta \left[ \frac{4}{3} + \frac{4(\varepsilon + E_F)}{9t_2} - k_y^2 \right] \nonumber \\
&~~~~~= \frac{W}{\pi} \sqrt{\left[\frac{4}{3} + \frac{4(\varepsilon + E_F)}{9t_2} \right].} 
\end{align}

For a transverse mode incident at the junction interface at an angle \( \theta \), the velocity matrix element $\hat{v}_y$ reads off  \cite{PhysRevB.98.075151,PhysRevB.94.094514,PhysRevB.101.214508}
\begin{equation}
    \hat{v}_y=\hat{v}_x\bigg(\tan \theta -\tan \theta'+ \frac{\delta y_{h}}{W}\bigg).
\end{equation}
 Note that $\theta\neq\theta'$ due to the quasi-flat nature of the band, as shown before. Note also, this quasi-flat band results in asymmetric Fermi contours in $k_x-k_y$ plane for different incident angles. Together these lead to finite $\hat{v}_y$ and the total current is obtained to be 
\begin{equation}
    \langle\mathfrak{J}_y\rangle=\frac{ \mathcal{N}_W }{WL}\sum_{k_y}~ \Theta~[\varepsilon(k_y) -E_F]~j_y.
\end{equation}
Fig.~\ref{fig:7}a shows the total current $\langle\mathfrak{J}_y\rangle$ along the propagation direction of Andreev modes  as a function of junction length  \( L \). We find an overall oscillatory nature of $\langle\mathfrak{J}_y\rangle$. This is due to quantum interference of the quasi-particle states reflecting between the NS interfaces. The superconducting wave functions penetrating the normal region acquire both decay and phase winding. This in turn leads to constructive or destructive interference depending on the junction length \cite{PhysRevB.49.498}. Figure~\ref{fig:7}b shows the variation of the $y$-directional current with respect to the incident angle ($\theta$) of the Andreev modes and the junction length ($L$). The current attains higher values when the incident angle approaches \(\pi/2\), corresponding to normal incidence. Moreover, in Fig.~\ref{fig:8}, we present the transverse current profile, which exhibits constant hall-like plateaus as a function of the phase difference. The nature of the plateaus are almost identical for different incident angles of electrons. This setup allows for the investigation of the planar Hall effect in Josephson currents by analyzing the response to a longitudinal phase bias and holds potential applications in Hall rectifiers\cite{PhysRevB.84.054520, PhysRevB.77.014528,PhysRevB.111.L020507,sahoo2025fourterminaljosephsonjunctionsdiode}.

\section{Conclusions}
In summary, we have theoretically investigated the effect of quasi-flat bands on Andreev reflection (AR) and related transport phenomena in both NS and SNS junction geometries  in a two-dimensional \(\alpha-\mathcal{T}_3\) lattice systems. By incorporating next-nearest-neighbor hopping, we demonstrate how tuning the band flatness leads to an improved AR, which in turn critically impacts the differential conductance of the junction. Our analysis shows that the flat band not only facilitates nearly perfect electron–hole conversion at the NS interface but also introduces pronounced directional asymmetries that manifest as sizable electronic GH shifts.
Furthermore, by tracking the real-time evolution of Gaussian wave packets, we shed light on the electron-hole conversion process in such systems. This dynamic behavior is essential for understanding the microscopic processes governing the interface scattering and explains the non-trivial spatial shifts observed in the reflected quasi-particle beams. In the context of Josephson physics, the analysis of Andreev bound states (ABS) in SNS junction led to a scaling behavior in the phase-dependent Josephson current even in regimes characterized by enhanced band flatness. Additionally, the expected oscillatory behavior of the critical Josephson current with junction length is preserved, and we observe the mean critical value of these currents can be tuned with band flatness. Moreover, the emergence of transverse charge currents due to the flat bands suggests that such systems can offer a suitable platform for engineering novel Josephson effects, with potential applications in superconducting waveguide devices \cite{kumar2025low,
zotova2024tunable} and Hall rectifiers \cite{PhysRevB.84.054520, PhysRevB.77.014528,PhysRevB.111.L020507,sahoo2025fourterminaljosephsonjunctionsdiode}. 

In sum, our results underline the significant role of flat bands in modifying the electronic scattering processes at metal-superconductor interfaces. The combination of enhanced AR probabilities, substantial GH shifts and modified Josephson transport enables the way of designing quantum devices \cite{valdes2025giant,
yang2025tunable} where the interplay between superconductivity and flat-band physics can be tuneable. Future experimental studies aimed at verifying these predictions could open up new avenues in electronics and contribute to ongoing efforts to exploit flat-band superconductivity in advanced material platforms\cite{PhysRevB.111.134502,
wu2025flat,
li2025electron,
guo2025superconductivity,
choi2025superconductivity}.

\section{Acknowledgments}
SM acknowledges the Virgo cluster, where most of the numerical calculations were performed. KS and AM acknowledge financial support from the Department of Atomic Energy (DAE), Govt. of India, through the project Basic Research in Physical and Multidisciplinary Sciences via RIN4001.

\appendix
\section{NS junction computations}
\label{AppendixA}
\subsection{Dirac-Bogoliubov-de Gennes equation}

The quasiparticle excitations in the NS junction can be expressed as \cite{PhysRevLett.97.067007,PhysRevLett.74.1657,PhysRevB.108.075425,RevModPhys.80.1337,de2018superconductivity}

\begin{equation}
\mathcal{\hat{H}}_{DBdG}\equiv
\begin{pmatrix}
\mathscr{H} - E_{\text{F}} & \mathbf{\Delta(x) \mathds{1}} \\
\mathbf{\Delta(x)} \mathds{1}^{\dagger} & E_{\text{F}} - \mathscr{THT}^{-1}
\end{pmatrix}
\begin{pmatrix}
u_{e} \\
v_{h}
\end{pmatrix}
= \varepsilon
\begin{pmatrix}
u_{e} \\
v_{h}
\end{pmatrix}.
\label{eq4}
\end{equation}
 Here, $\mathscr{H}=\diag[\hat{\mathcal{H}}^{+}(\mathbf{k}),\hat{\mathcal{H}}^{-}(\mathbf{k})]$, spanned over two valleys ($K,K'$). The components $u_{e}~ (v_{h})$ represent the electron (hole) components of quasi-particle eigenstates, with the excitation energy $\varepsilon$ measured from the Fermi Energy ($E_F$). The time reversal operator, responsible for switching between valleys, is denoted as $\mathscr{T}$, defined by$\mathscr{T}=\sigma_x\otimes\tau\mathcal{K}$, where $\sigma_x$ represents the Pauli matrix, $\tau$ acts on valley subspaces, and $\mathcal{K}$ is the operator for complex conjugation \cite{PhysRevLett.97.067007,PhysRevB.104.125441}. The Hamiltonian for $\alpha-\mathcal{T}_3$ preserves time-reversal symmetry (TRS), ensuring $\mathscr{THT}^{-1}=\mathscr{H}$. Hence, the DBdG Hamiltonian, as described in Eq.~\ref{eq4}, which typically comprises a $12\times12$ matrix, due to TRS, can be rewritten as
 
\begin{equation}
\begin{pmatrix}
\hat{\mathcal{H}}^{\pm}(\mathbf{k}) - E_{\text{F}} & \mathbf{\Delta(x) \mathds{1}} \\
\mathbf{\Delta(x)} \mathds{1}^{\dagger} & E_{\text{F}} - \hat{\mathcal{H}}^{\pm}(\mathbf{k})
\end{pmatrix}
\begin{pmatrix}
u_{e} \\
v_{h}
\end{pmatrix}
= \varepsilon
\begin{pmatrix}
u_{e} \\
v_{h}
\end{pmatrix}.
\label{eq5}
\end{equation}
It effectively reduces it to two sets of $6$ decoupled equations due to the valley degeneracy and we will continue our calculation by taking $\hat{\mathcal{H}}^{+}(\mathbf{k})$ for convenience \cite{PhysRevB.104.125441,PhysRevB.108.085423,PhysRevB.106.245106,PhysRevLett.97.067007}.

The decoupled DBdg equation can be represented as

\begin{widetext}
\begin{equation}
\resizebox{1\textwidth}{!}{%
$\begin{bmatrix}
2 t_2 \xi(\mathbf{k})-E_F & f_{\mathbf{k}}^{+} \cos \varphi & 0 & \mathbf{\Delta(x)} & 0 & 0 \\
f_{\mathbf{k}}^{*+} \cos \varphi & 2 t_2 \xi(\mathbf{k})-E_F & f_{\mathbf{k}}^{+} \sin \varphi & 0 & \mathbf{\Delta(x)} & 0 \\
0 & f_{\mathbf{k}}^{*+} \sin \varphi & 2 t_2 \xi(\mathbf{k})-E_F & 0 & 0 & \mathbf{\Delta(x)} \\
\mathbf{\Delta(x)}^{*} & 0 & 0 & E_F-2 t_2 \xi(\mathbf{k}) & -f_{\mathbf{k}}^{+} \cos \varphi & 0 \\
0 & \mathbf{\Delta(x)}^{*} & 0 & -f_{\mathbf{k}}^{*+} \cos \varphi &  E_F-2 t_2 \xi(\mathbf{k}) & -f_{\mathbf{k}}^{+} \sin \varphi \\
0 & 0 & \mathbf{\Delta(x)}^{*} & 0 & -f_{\mathbf{k}}^{*+} \sin \varphi &  E_F-2 t_2 \xi(\mathbf{k})
\end{bmatrix}
\begin{bmatrix}
u_1 \\
u_2 \\
u_3 \\
v_1 \\
v_2 \\
v_3
\end{bmatrix}
=
\varepsilon
\begin{bmatrix}
u_1 \\
u_2 \\
u_3 \\
v_1 \\
v_2 \\
v_3.
\end{bmatrix}$}
\label{eq6}
\end{equation}
\end{widetext}
The dispersion obtained from Eq.~\ref{eq6} encompasses six branches (three for electrons and three for holes), but our focus will be solely on the two branches corresponding to quasi-flat bands. It is also important to consider a gate voltage potential $U(x) = -U_0\Theta(x)$ which can be adjusted by doping in the superconducting region and is zero in the normal region. The condition for the validity of the mean-field condition in the superconducting region is $|\mathbf{\Delta(x)}|\ll (E_F+U_0)$\cite{PhysRevLett.97.067007,PhysRevB.104.125441,Islam_2022,Puglia_2021, Nakagawa_2021}. The quasi-flat band spectra of the total system are given by,
\begin{equation}
\varepsilon=\pm\sqrt{(E_F+U(x)+2 t_2 \xi(\mathbf{k}))^2+\left|\mathbf{\Delta(x)}\right|^2\Theta(x)}
    \label{eq7}
\end{equation}
For the reflection upon the junction interface ($x=0$), the energy $\varepsilon$ and transverse wave vector $k_y$ are conserved. This gives us the eigenfunctions in the normal region in plane wave basis as

\begin{equation}
    \psi_{e}^{\pm} = e^{\left(\pm i k^{e}_{x} x + i k_{y} y\right)}
\begin{pmatrix}
\pm e^{\mp i \theta} \cos \varphi \\
1 \\
\pm e^{\pm i \theta} \sin \varphi \\
0 \\
0 \\
0
\end{pmatrix}
\label{eq8}
\end{equation}
and for the hole-like part

\begin{equation}
\psi_{h}^{\pm} = e^{\left(\pm i k^h_{x} x + i k_{y} y\right)}
\begin{pmatrix}
0 \\
0 \\
0 \\
\mp e^{\mp i \theta^{\prime}} \cos \varphi \\
1 \\
\mp e^{\pm i \theta^{\prime}} \sin \varphi
\end{pmatrix}.
\label{eq9}
\end{equation}

The superscript $(\pm)$ on the wave functions denotes the direction of propagation of the electron (hole) towards (away) from the junction interface. Here the longitudinal wave vector for the electron (hole) can be represented as \begin{equation}
    k^h_{x}(k^e_x)=\sqrt{\left[\frac{4}{3}+\frac{4( E_F-(+)\varepsilon)}{9t_2}\right]-k_y^2}
    \label{eq10}
\end{equation}
and the incident angle of the electron and the reflected angle of the hole is denoted by $\theta=\sin^{-1}[\hbar v_{F}k_{y}/(\varepsilon+E_{\rm F}-2 t_2 \xi(\mathbf{k}))]$ and $\theta^{'}=\sin^{-1}[\hbar v_{F}k_{y}/(\varepsilon-E_{\rm F}+2 t_2 \xi(\mathbf{k}))]$ respectively. Using Eq.~\ref{eq7}, we find out the critical angle ($\theta_c$) for AR to be $\theta_c=\sin^{-1}\left[\left|\dfrac{\varepsilon-E_{\rm F}+2 t_2 \xi(\mathbf{k})}{\varepsilon+E_{\rm F}-2 t_2 \xi(\mathbf{k})}\right|\right]$. For $|\theta|>\theta_c$ AR is no longer allowed. With this, the wave function  in the superconducting region can be expressed as

\begin{eqnarray}
\psi^{\pm}_{\rm S}=\begin{pmatrix}e^{\pm i\beta}\\\pm\frac{1}{\cos\varphi}e^{\pm i\beta}\\\tan\varphi e^{\pm i\beta}\\e^{-i\phi}\\\pm\frac{1}{\cos\varphi}e^{-i\phi}\\\tan\varphi e^{-i\phi}\end{pmatrix}e^{\pm ik_{0}x+ik_{y}y-\kappa x},
\label{eq11}
\end{eqnarray}
where $k_{0}=U_{0}/\hbar v_{F}$, $\kappa=(\Delta_{0}/\hbar v_{F})/\sin\beta$, and $\beta$ is defined as
\begin{equation}
\beta=
\begin{cases}
\cos^{-1}(\varepsilon/\Delta_{0})&\text{$\varepsilon<\Delta_{0}$},\\
-i\cosh^{-1}(\varepsilon/\Delta_{0})&\text{$\varepsilon>\Delta_{0}$}.
\end{cases}
\label{eq12}
\end{equation}
The state $\psi_{\mathrm{S}}^{+}\left(\psi_{\mathrm{S}}^{-}\right)$ denotes the wave function of a quasi-hole (quasi-electron) when $\varepsilon>\Delta_{0}$. Conversely, within the superconducting region, for $\varepsilon<\Delta_{0}$, this state represents the coherent superposition of electron and hole excitations. Considering the conservation of probability current in $x$-direction, we follow the derivations of probability current $\mathbf{\mathcal{J}}$ utilizing the continuity equation by assuming a wave function in the general form $\Phi=\left(\psi_{A}, \psi_{B}, \psi_{C}\right)^{\mathrm{T}}$, which satisfies $\partial_{t} \left|\Phi\right|^2+\nabla\cdot\mathbf{\mathcal{J}}=0$. This allows us to match the boundary condition at the junction interface by which we can derive the electron reflection probability ($R$) and Andreev reflected probability ($R_A$)
\cite{PhysRevB.102.045132,PhysRevLett.97.067007,PhysRevB.106.094503,PhysRevB.104.125441,Islam_2022}.  

\begin{figure}
    \centering
    \includegraphics[width =0.4\textwidth]{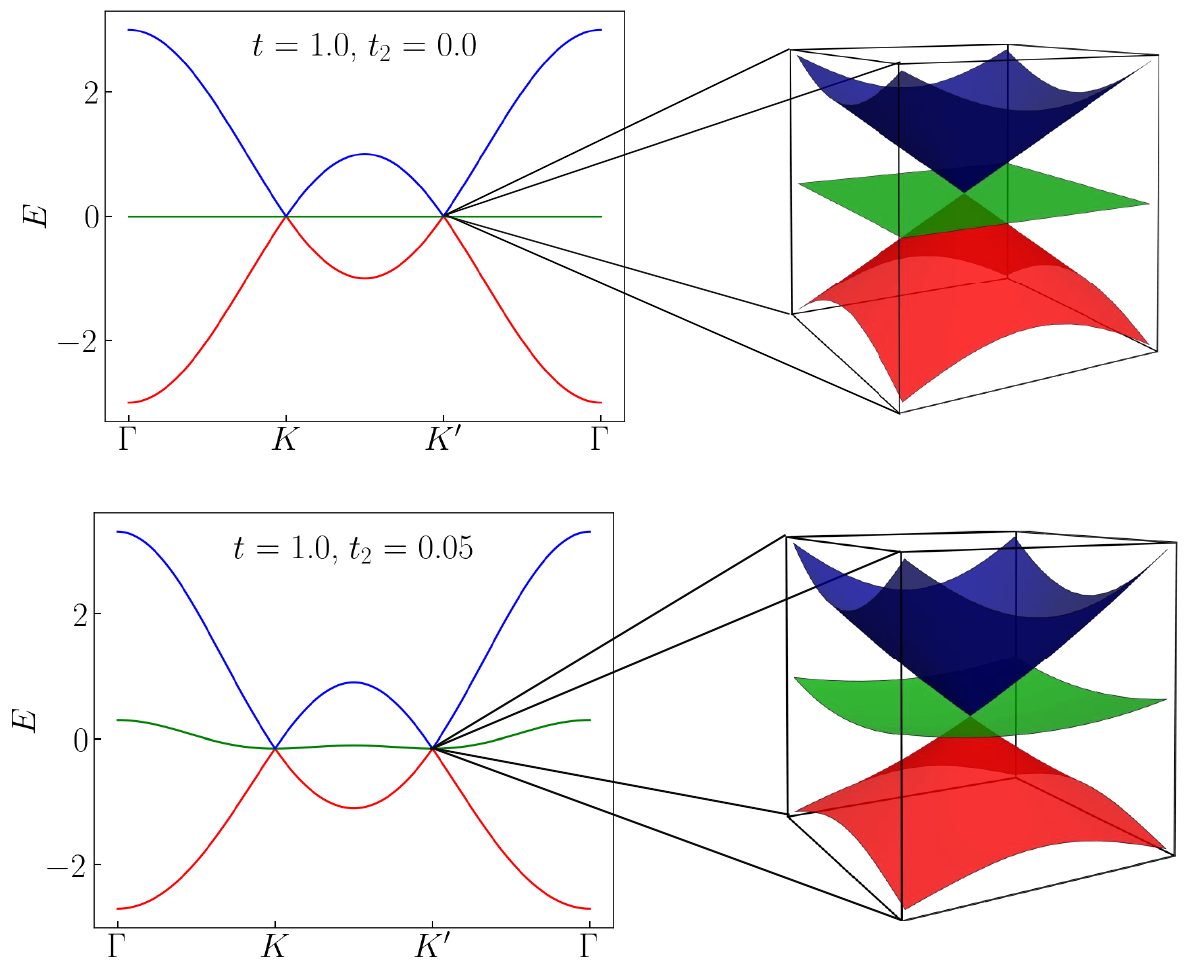}
    \vspace{-1\baselineskip}
    \caption{\textbf{ Plot of tight-binding dispersion: } The upper panel shows the tight-binding band structure along the highly symmetric point of the Brillouin zone for $\alpha-\mathcal{T}_3$ lattice. On the right, we see the low energy dispersion around one of the Dirac points ($K, K'$). The dispersion is linear and the flat band is present at $\epsilon = 0$, shown in green. In the presence of NNN hopping $t_2=0.05$, the flat band becomes nearly dispersive, shown in green colour on the lower panel.} 
    \label{fig:A1}
\end{figure}

\begin{figure}
    \centering
    \includegraphics[width =0.5\textwidth]{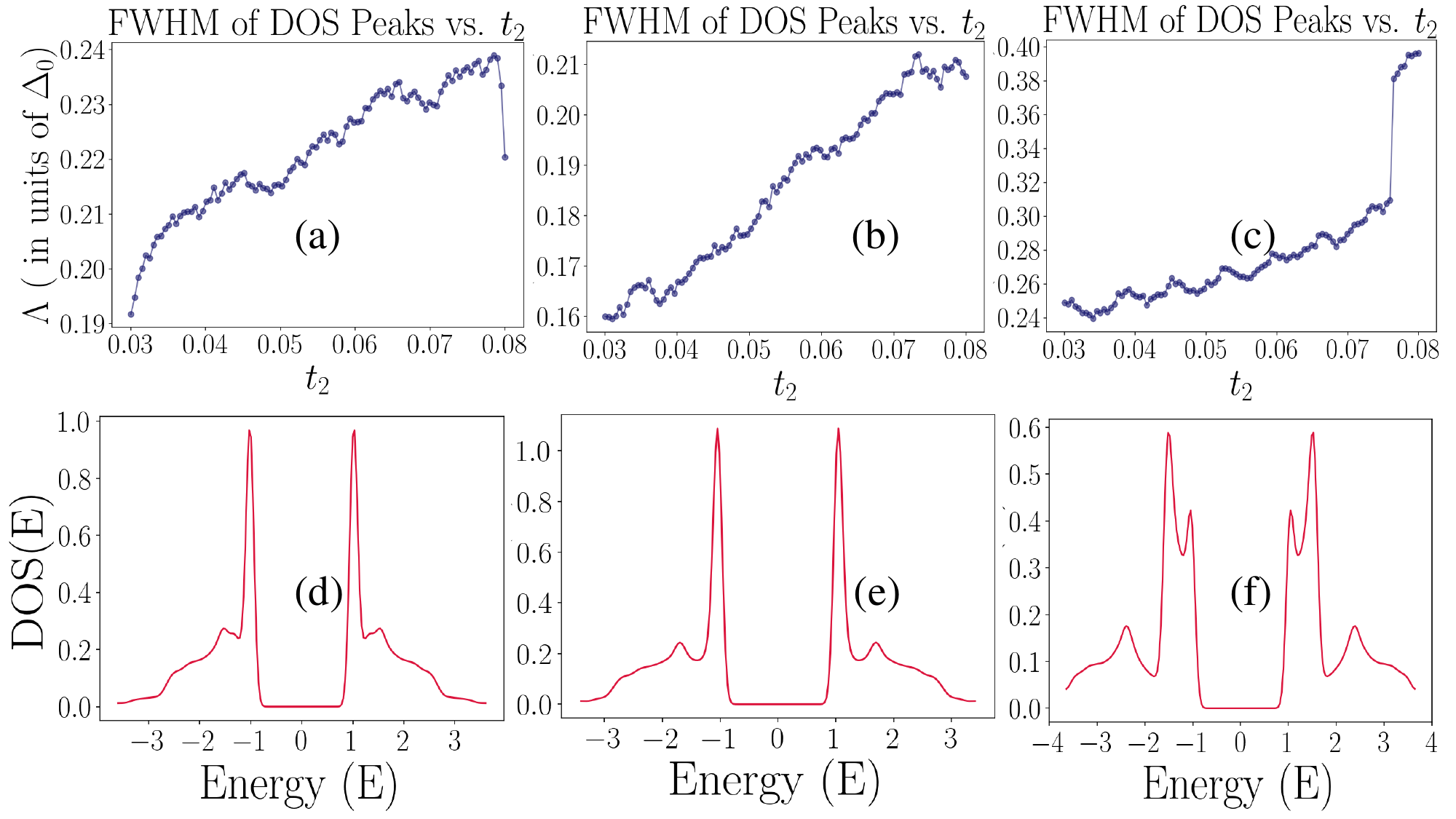}
    \vspace{-1\baselineskip}
    \caption{\textbf{ Density of States \& FWHM:} In panel (a-c) the variation of FWHM of the DOS (d-f) peaks has been shown with respect to the band flatness tuning parameter $t_2$. For both upper and lower panel the Fermi Energy is set to be $E_F=0,~0.2\Delta_0$ \& $E_F=\Delta_0$ respectively.  }
    \label{fig:A2}
\end{figure}

\section{Density of States \& Bandwidths}
\label{appenB}

The decoupled Bogoliubov–de Gennes (BdG) equation, represented in Eq. (\ref{eq6}), describes the system under consideration. The density of states (DOS) of this system exhibits prominent peaks at energies corresponding to the superconducting gap parameter, $( E = \Delta_0 )$. These peaks arise from the quasi-flat nature of the bands, which are characteristic of the system. However, the bandwidths of these DOS peaks depend on the degree of flatness of the bands, which is controlled by the parameter \( t_2 \). To analyze this numerically, we compute the eigenvalues of the Hamiltonian for varying values of \( t_2 \) over a discretized Brillouin zone defined by \( k_x \) and \( k_y \). The eigenvalues are used to construct a histogram representing the DOS, which is then smoothed using a Gaussian filter to suppress numerical noise. The smoothed DOS is analyzed to identify peaks, and their bandwidths are quantified by calculating the full width at half-maximum (FWHM) of each peak denoted by $\Lambda$. In Fig.~\ref{fig:A2} , we present the variation of the bandwidths as a function of \( t_2 \) for different Fermi energy values. The results demonstrate that as \( t_2 \) increases, the DOS peaks broaden, reflecting the transition of the bands from being highly flat to progressively more dispersive. The information on the bandwidths gives us a systematic way to vary the incident electron energy ($\varepsilon$) and the Fermi energy ($E_F$) in the vicinity of the flat band and more precisely with the range of $E_F \in [\Delta_0-\frac{\Lambda}{2},\Delta_0+\frac{\Lambda}{2}]$. Also to ensure internal consistency between the energy dispersion and momentum-dependent parameters, we compute the quantity $\xi(\varepsilon, \theta)$ self-consistently using a fixed-point iteration scheme. Starting from an initial guess $\xi^{(0)} = 0.1$, we impose the constraint that the total wave-vector magnitude $k^2 = k_x^2 + k_y^2$ must satisfy the energy dispersion relation $\varepsilon = \varepsilon(\mathbf{k})$. At each iteration, we compute $k_y = \sqrt{k^2} \sin\theta$, and deduce $k_x^2 = k^2 - k_y^2$, then update $\xi$ via $\xi^{(n+1)}(\mathbf{k}) = \xi(\mathbf{k})$. The iteration proceeds until convergence within a tolerance of $10^{-6}$, ensuring that $\xi$ is consistent with the underlying band structure and momentum components for each angle and excitation energy.

\begin{widetext}
\section{Andreev Bound State Spectra}
\label{appenC}
To compute the ABS spectra mentioned in Section~\ref{sns}, we solve a set of eight linear equations written compactly as $\mathbb{A} \mathbb{X} = \mathbb{0}$, where $\mathbb{A}$ is an $8 \times 8$ coefficient matrix determined by boundary conditions and wavefunction matching at the NS interfaces. For nontrivial solutions to exist (i.e., $\mathbb{X} \neq \mathbb{0}$), the determinant of the matrix must vanish. Therefore, we compute the ABS spectra by solving $\det \mathbb{A} = 0$, which yields the allowed energy eigenvalues corresponding to bound states in the junction.
\begin{align}
\det\mathbb{A}=\mathbf{\mathscr{F}}(\varepsilon,E_F,L,\theta,\phi,\alpha)
&= -2 \bigg(-1 - 2 \alpha^2 - 4 \alpha^4 - 2 \alpha^6 - \alpha^8 
+ 2 (1 + \alpha^2 + \alpha^4)^2 \varepsilon^2 \nonumber\\
& - \alpha^2 (1 + \alpha^4 - 2 (1 + \alpha^2 + \alpha^4) \varepsilon^2) \cos(2\theta) \nonumber\\
& - \alpha^2 \big(1 + \alpha^4 - 2 (1 + \alpha^2 + \alpha^4) \varepsilon^2 
- 2 \alpha^2 (-1 + \varepsilon^2) \cos(2\theta) \big) \cos(2\theta') \nonumber\\
& + (-1 + \alpha^4)^2 \sin(\theta) \sin(\theta') \bigg) 
\sin^2(E_F L \cos(\theta)) \nonumber\\
& - 2 \alpha^2 (1 + \alpha^2)^2 \varepsilon \sqrt{1 - \varepsilon^2} 
\cos^2(\theta) \cos(\theta') \sin(2 E_F L \cos(\theta)) \nonumber\\
& - (1 + \alpha^2)^2 \varepsilon \sqrt{1 - \varepsilon^2} 
(2 + \alpha^2 + 2 \alpha^4 + \alpha^2 \cos(2\theta)) 
\cos(\theta') \sin(2 E_F L \cos(\theta)) \nonumber\\
& + (1 + \alpha^2)^2 \cos(\theta) \bigg( 
2 (1 + \alpha^2)^2 \cos(\theta') (\cos(\phi) 
+ (1 - 2 \varepsilon^2) \cos^2(E_F L \cos(\theta)))\nonumber\\
& + 2 \alpha^2 \varepsilon \sqrt{1 - \varepsilon^2} \cos^2(\theta') \sin(2 E_F L \cos(\theta)) \nonumber\\
& + \varepsilon \sqrt{1 - \varepsilon^2} (2 + \alpha^2 + 2 \alpha^4 + \alpha^2 \cos(2\theta')) \sin(2 E_F L \cos(\theta)) 
\bigg)
\label{eqa1}
\end{align}
\begin{figure}
    \centering
    \includegraphics[width =0.4\textwidth]{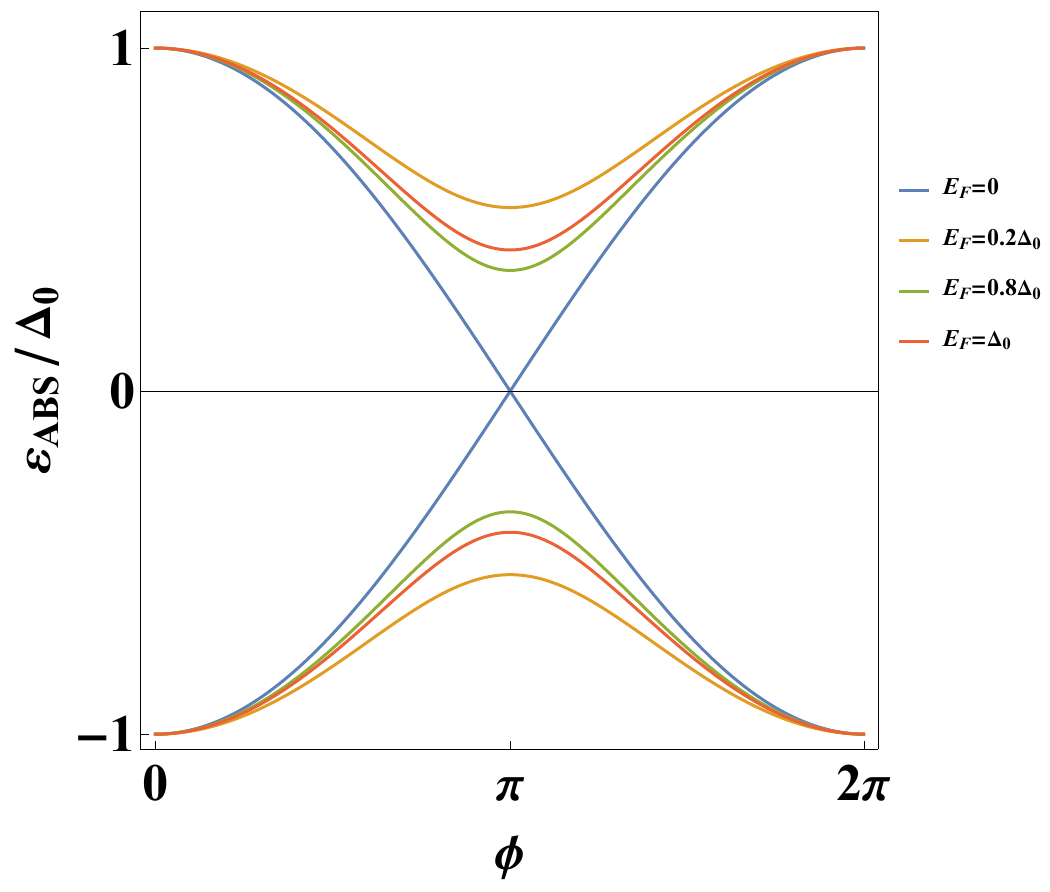}
    \vspace{-1\baselineskip}
    \caption{\textbf{ ABS spectra :} The ABS spectra for different $E_F$ values for parameters $L=100$ nm, $\theta=\pi/4$.}
    \label{fig:A3}
\end{figure}
We solve Eq.~\ref{eqa1} for $\det\mathbb{A}=0$, and numerically we get the ABS spectra shown in Fig.~\ref{fig:A3} for different $E_F$.

\end{widetext}

\bibliography{references}

\begin{thebibliography}{99}
\bibitem [{dis()}]{discussion}%
  \BibitemOpen
  \href@noop {} {}\bibinfo {note} {The superconducting coherence length, when transformed into MKS units is $\sim 658.2$ nm, hence we always have the junction length lower than the superconducting coherence length in our computations.}\BibitemShut {Stop}%
\end{thebibliography}
\end{document}